\documentclass[10pt, journal]{IEEEtran}

\usepackage{times}
\usepackage{graphicx,subfigure,epsfig,overpic,color}
\usepackage{url,amsmath,amssymb,bm}
\usepackage{multirow}
\usepackage{float}
\usepackage{booktabs}
\usepackage{threeparttable,array,tabularx,extarrows}

\usepackage[ruled,boxed]{algorithm2e}
\usepackage{lineno}
\usepackage{epstopdf}

\ifCLASSOPTIONcompsoc
\else
\fi

\ifCLASSINFOpdf
\else
\fi

\begin{document}

\title{Finding A Taxi with Illegal Driver Substitution Activity via Behavior Modelings} 
%
%
\author{Junbiao~Pang, Muhammad Ayub Sabir, Zhuyun~Wang, Anjing~Hu, Xue~Yang, Haitao~Yu, and Qingming~Huang
\IEEEcompsocitemizethanks{

\IEEEcompsocthanksitem J. Pang, M. Sabir and A. Hu are with the Faculty of Information Technology, Beijing University of Technology, Beijing 100124, China (e-mail: \mbox{junbiao\_pang@bjut.edu.cn}).

\IEEEcompsocthanksitem  Z. Wang is the Beijing Municipal Transportation Law Enforcement Corps, Beijing 100044, China.

\IEEEcompsocthanksitem  X. Yang and H. Yu are with the Beijing Transportation Information Center,
Beijing 100161, China.

\IEEEcompsocthanksitem  Q. Huang is with the University of Chinese Academy of Sciences, Chinese Academy of Sciences (CAS), Beijing
100049, China, and the Institute of Computing Technology, CAS, Beijing
100190, China.
 }
}

\markboth{Submitted to }%
{Shell \MakeLowercase{\textit{et al.}}:Finding A Taxi with Illegal Driver Substitution Activity via Driver Behavior Modelings}

\IEEEcompsoctitleabstractindextext{%

\begin{abstract}

In our urban life, Illegal Driver Substitution (IDS) activity for a taxi is a grave unlawful activity in the taxi industry, possibly causing severe traffic accidents and painful social repercussions. Currently, the IDS activity is manually supervised by law enforcers, i.e., law enforcers empirically choose a taxi and inspect it. The pressing problem of this scheme is the dilemma between the limited number of law-enforcers and the large volume of taxis. In this paper, motivated by this problem, we propose a computational method that helps law enforcers efficiently find the taxis which tend to have the IDS activity. Firstly, our method converts the identification of the IDS activity to a supervised learning task. Secondly, two kinds of taxi driver behaviors, i.e., the Sleeping Time and Location (STL) behavior and the Pick-Up (PU) behavior are proposed. Thirdly, the multiple scale pooling on self-similarity is proposed to encode the individual behaviors into the universal features for all taxis. Finally, a Multiple Component- Multiple Instance Learning (MC-MIL) method is proposed to handle the deficiency of the behavior features and to align the behavior features simultaneously. Extensive experiments on a real-world data set shows that the proposed behavior features have a good generalization ability across different classifiers, and the proposed MC-MIL method suppresses the baseline methods.

\end{abstract}

}

\maketitle

\IEEEdisplaynotcompsoctitleabstractindextext
\IEEEpeerreviewmaketitle

\section{Introduction}

\newtheorem{myobr}{Observation}
\newtheorem{mydef}{Definition}
\newtheorem{mythe}{Theorem}
\newtheorem{mypro}{Proposition}

Taxis play a pivotal role in urban transportation, offering dynamic, convenient, and time-efficient door-to-door services. Yet, managing taxi operations presents unique challenges due to their highly flexible routes, passenger demands, and operational hours, distinguishing them from more static transportation systems like buses and subways. A notable issue within the taxi industry is Illegal Driver Substitution (IDS), where:

\begin{mydef}[Definition of IDS]
A taxi is operated by someone other than the legally registered driver, in violation of their contractual agreement.
\end{mydef}

Local regulations stipulate that a legal taxi driver must possess a vocational license and a formal contract with a taxi company. IDS can manifest in two primary forms:
\begin{itemize}
\item Operation of a taxi by an individual lacking the necessary vocational license;
\item Operation by a licensed driver who is not officially registered to the vehicle in question.
\end{itemize}
Thus, IDS occurs whenever a taxi is used by unauthorized personnel to provide transportation services, often motivated by the illicit objective of maximizing revenue through the subcontracting of taxi services to unregistered individuals.
This illicit practice poses significant risks, including severe traffic incidents and criminal acts (e.g., robbery, murder), undermining the safety and integrity of the taxi industry and local governance. 


Currently, law enforcement's manual checks on taxis are insufficient due to the sheer volume of taxis compared to the limited number of inspectors. This situation prompts a vital question for transportation safety officials: how can we more effectively identify taxis engaged in IDS?

\begin{figure*}[t!]
  \centering
  \subfigure[]{
    \label{fig:subfig:age} 
    \includegraphics[width=.3\textwidth]{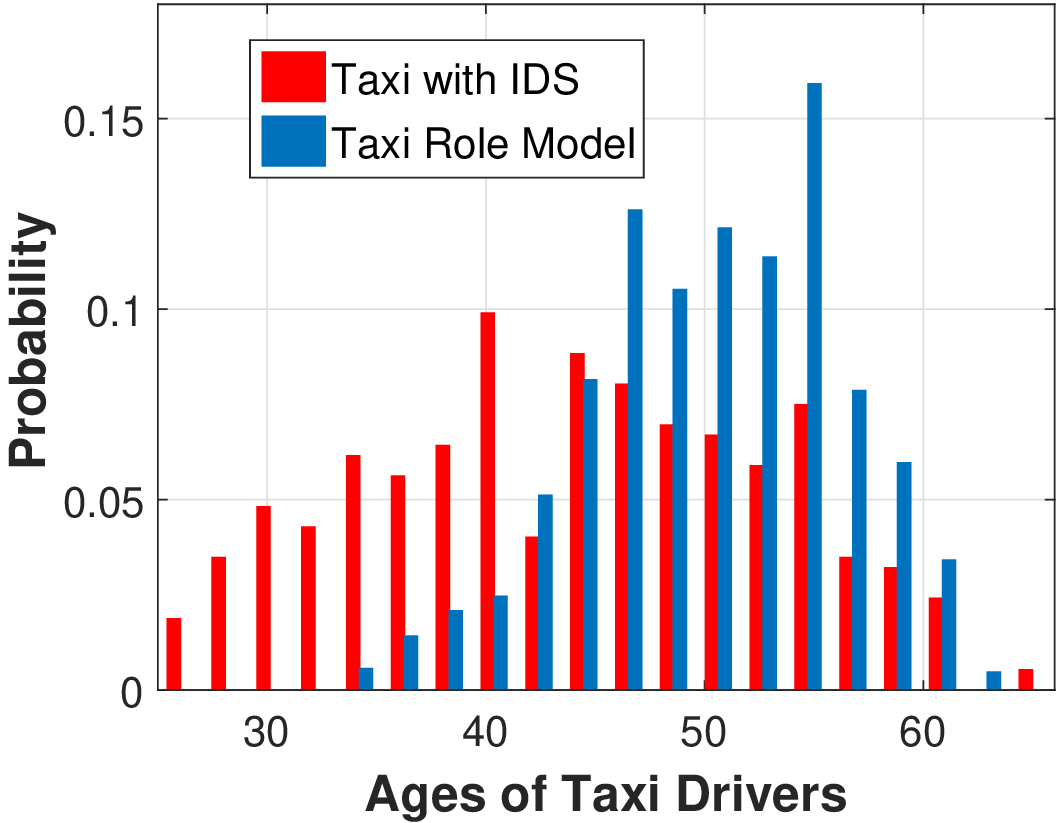}}
    \hspace{.1cm}
  \subfigure[]{
    \label{fig:subfig:eduaction} 
    \includegraphics[width=.3\textwidth]{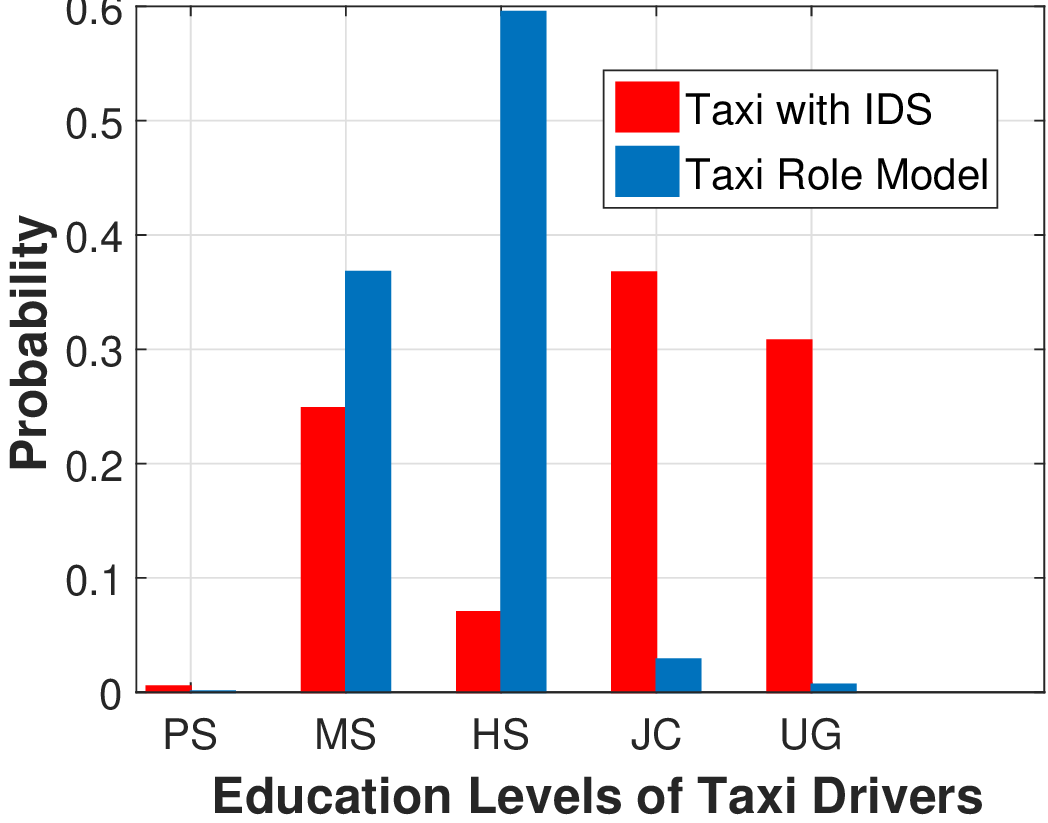}}
      \hspace{.1cm}
  \subfigure[]{
    \label{fig:subfig:year_taxi_driver} 
    \includegraphics[width=.3\textwidth]{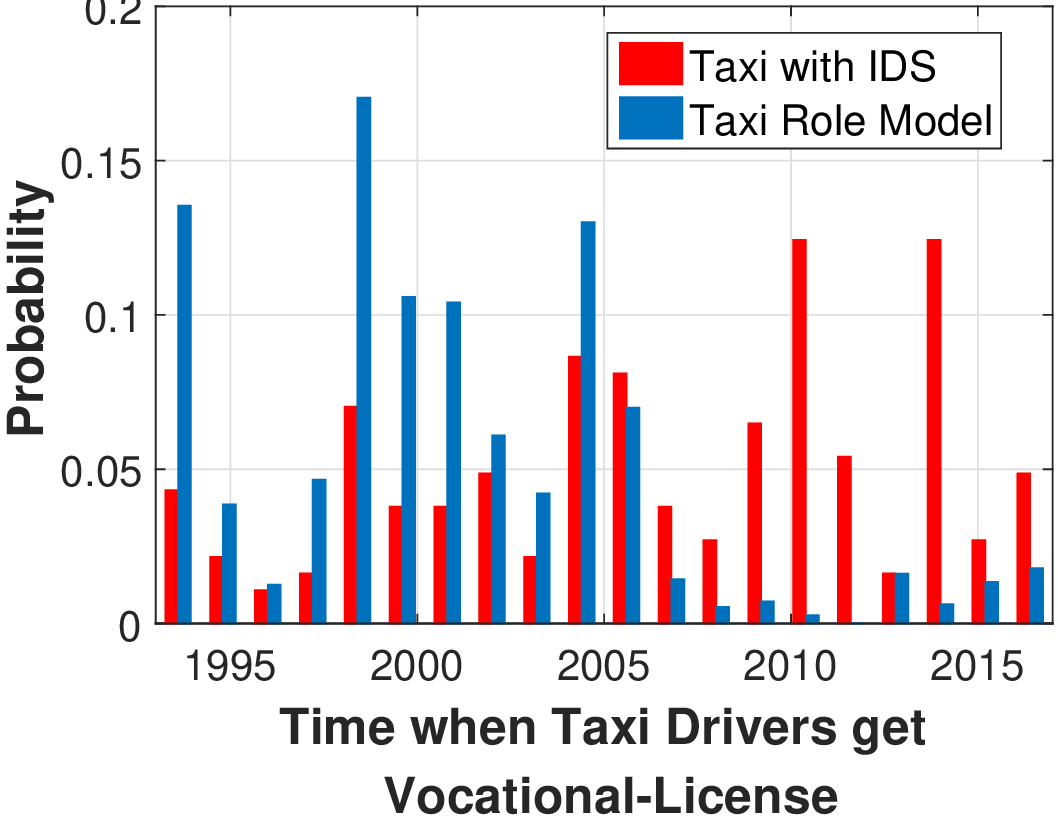}}
  \caption{Unreliable driver profile (best viewed in color). (a) The comparison of the ages between taxis with IDS and taxi role models. (b) The comparison of the education level between taxis with IDS and taxi role models; ``PS'', ``MS'', ``HS'', ``JC'', and ``UG'' are the abbreviation of Primary School, Middle School, High School, Junior College, and UnderGraduate, respectively. (c) The comparison of the first time when a taxi obtained the vocational license.}
  \label{fig:taxi_profile} 
\end{figure*}

A promising strategy involves leveraging Global Positioning Satellite (GPS) data, taximeter records, and driver registration information. GPS data can reveal driver behavior patterns, such as resting habits or meal times \cite{zhang2014understanding} \cite{xu2016taxi}, while taximeter records offer insights into service income and passenger pickup/drop-off locations \cite{pang2017discovering} \cite{zhang2016inferring}. Additionally, driver profiles (e.g., education, experience) could aid in identifying IDS activities, given the variability and inconsistency of IDS events among different drivers. This variability highlights the importance of monitoring for changes in driver behavior as a key indicator of IDS.

To unlock the potential of taxi service data, it's crucial to transform individual driver behaviors into \emph{Common} and \emph{Consistent} features applicable across the fleet. This entails:
\begin{itemize}
\item []\textbf{Commonness.} GPS data and taximeter records reflect individual behaviors, but for IDS analysis, extracting shared patterns across taxis is more valuable. This approach simplifies classifier design and enhances our understanding of IDS by focusing on commonalities rather than individual variances.
\item []\textbf{Consistency.} The variability in driver data, due to inaccuracies or deceptive practices, necessitates a focus on consistent behaviors for feature design. This ensures our models are built on reliable data, crucial for detecting IDS activities accurately.
\end{itemize}

This paper delves into the challenge of identifying taxis engaging in IDS activities through behavioral modeling. Understanding the research challenges is a key first step in this exploration

\subsection{Research Challenges}

\textbf{Unreliable Driver Profile.} Driver-supplied profiles to taxi companies poorly predict involvement in IDS activities. Analysis depicted in Fig.\ref{fig:taxi_profile} shows minimal correlation between the profiles of IDS-engaged taxis and "taxi role models." Notably, attributes such as education, age, and timing of vocational licensing offer no reliable indicators of IDS participation. Research suggests that the key determinants of crime-age profiles are more closely associated with personal circumstances, including living arrangements, family interactions with law enforcement, and truancy rates\cite{hansen2003education}.

\textbf{Misaligned Behaviors.} The driving patterns among taxi drivers vary significantly. For instance, some drivers exhibit a preference for nocturnal shifts, whereas others opt for daytime hours. This diversity is underscored by a study analyzing taxi traces in Beijing, which found substantial variability in driver behavior~\cite{yuan2010t}.

\textbf{Data Imbalance.} The incidence of IDS within the taxi community, while serious, is infrequent. A focused study revealed that only a fraction (0.19
\subsection{Our Contribution}

To address these challenges, we propose two efficient and effective driver behaviors from the traces of taxis and the records of taximeters. Our technique takes advantage of the following two critical observations:
\begin{itemize}
\item [](\uppercase\expandafter{\romannumeral1})~\emph{Compared with the registered driver, the illegal one tends to have a different sleeping pattern.} Especially for the one-shift taxis, a driver tends to have a consistent sleeping behavior since a person usually has a relatively fixed domicile in a city. Besides, the sleeping pattern (\emph{e.g.}, the duration time of sleeping) tends to be different for different drivers.
\item [](\uppercase\expandafter{\romannumeral2})~\emph{The operating schemes (\emph{e.g.}, patterns from the distributions of PUs or DOs) between taxi drivers reflect their different driving behaviors.} Both empirical and social behavior studies (\emph{e.g.},~\cite{fazio1981direct}) demonstrate that, over a sufficiently long period of time, a person's behavior exhibits a surprisingly high level of consistency. Moreover, the spatio-temporal distribution of the PU points discovers the individual profit-hunting scheme, which typically reflects the behaviors of different drivers~\cite{zhang2016inferring}.
\end{itemize}

Utilizing insights from Observation (\uppercase\expandafter{\romannumeral1}), we model the Sleeping Time and Location (STL) of a taxi driver via Fisher Vector (FV)~\cite{sanchez2013image} from the following information: a) the GPS locations of a driver's sleeping locations, b) the start working time and the sleeping time. Based on (\uppercase\expandafter{\romannumeral2}), we utilize Latent Dirichilet Allocation (LDA)~\cite{blei2003latent} to learn the spatial-temporal PU behavior. Therefore, based on both (\uppercase\expandafter{\romannumeral 1}) and (\uppercase\expandafter{\romannumeral 2}), we propose two kinds of individual-wise driver behaviors.

The individual behaviors are the personal description of each driver. To efficiently discover the taxi with the IDS activity, the individual-wise driver behaviors should be further encoded into the discriminative features for all taxis. This paper proposes to combine the Self-Similarity (SS) approach and the pooling to discover the taxis with the IDS activity. To align these SS-based features over a long-time range, we propose Multiple Component-Multiple Instance Learning to handle the possibility of deficiency of the behavior features and align the behavior features.




This paper makes several significant contributions to the field, as outlined below:

\begin{itemize}
\item []\emph{1. Introduction to the IDS Problem.} This study is pioneering in exploring the detection of one-shift taxis engaged in IDS activities, utilizing GPS data and taximeter records. We approach this challenge as a supervised learning problem, achieving effective solutions for identifying taxis with the IDS activities.

\item []\emph{2. Modeling Driver Behavior.} We have developed the STL and PU behaviors, offering the insights into drivers' rest patterns and profitability strategies, respectively. These are instrumental in characterizing the nuanced activities of each driver.

\item []\emph{3. Multi-Scale Pooling (MSP) on Self-Similarity (SS).} Acknowledging the variable timing of the IDS activities among taxis, we introduce SS to pinpoint the IDS occurrences and MSP to standardize these into a uniform feature vector dimension. The synergy of SS and MSP effectively translates diverse individual driver behaviors into a common feature space applicable across all taxis.

\item []\emph{4. Multiple Component-Multiple Instance Learning (MC-MIL).} MC-MIL addresses the challenges of behavior feature deficiencies and aligns IDS-related features. Additionally, we evaluate the performance of deep learning based method (i.e., Long short-term memory (LSTM) and Transformer) on the imbalanced and small-scale dataset. Our findings reveal MC-MIL as the superior method among these classifiers for detecting IDS activities.
\end{itemize}

\begin{figure}[t!]
\centering
\includegraphics[width=.32\textwidth]{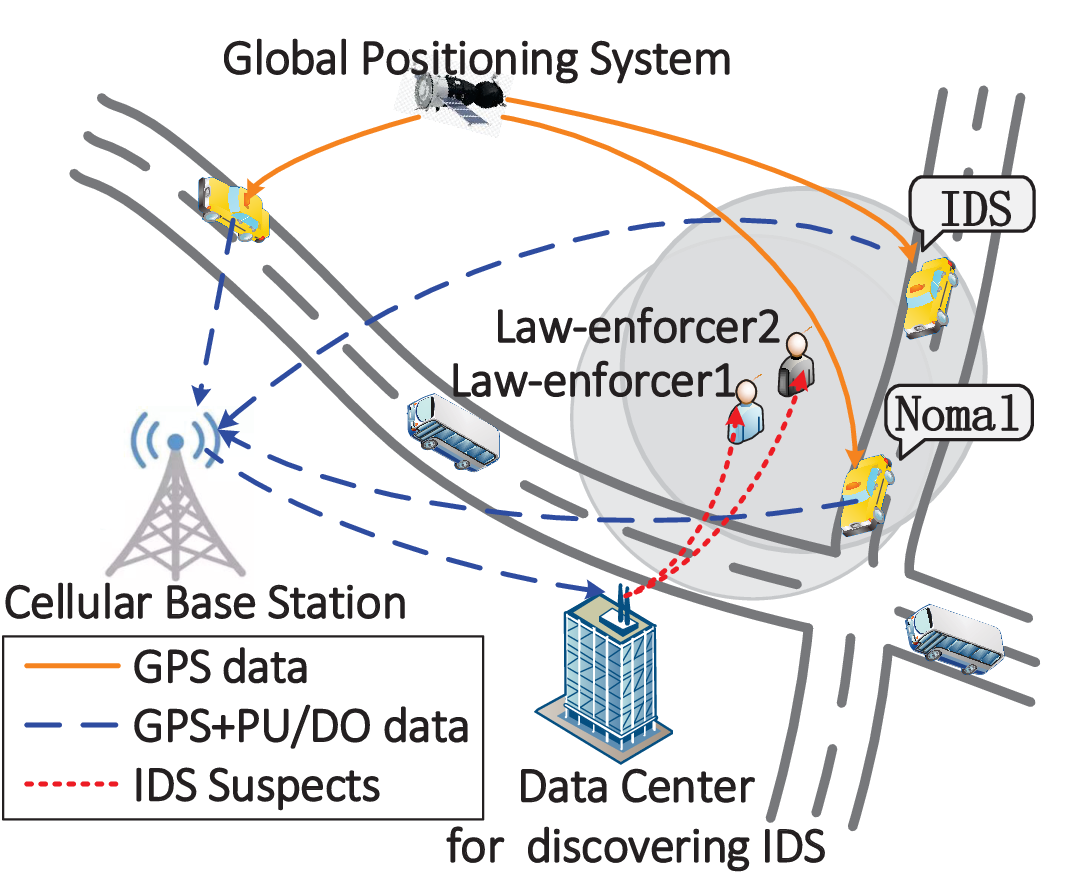}
\caption{The system of inspecting a taxi with the IDS activity for law-enforcers (best viewed in color).}\label{fig:SDA-framework}
\end{figure}


\section{Background and Related Work}\label{sec:relatedwork}

\subsection{Background}


As depicted in Fig.\ref{fig:SDA-framework}, leveraging GPS and PU/DO data allows a data center to identify and notify law enforcers of taxis engaging in IDS activities directly on their mobile devices. By setting a specific inspection area on their devices, law enforcers can strategically locate and manually verify suspect taxis through the electronic map, bypassing the need for random checks. Thus, the efficacy of the system in Fig.\ref{fig:SDA-framework} hinges on its ability to accurately and swiftly pinpoint taxis with IDS activities.

\subsection{Related Work}

\textbf{Pattern Analysis from GPS Traces:} A variety of research issues have been addressed by leveraging large-scale GPS traces, \emph{e.g.}, urban human mobility understanding~\cite{cao2005mining}~\cite{wang2017human}~\cite{pang2017discovering}, urban planning~\cite{yuan2011driving}~\cite{zheng2011urban}, traffic prediction~\cite{zhang2017analyzing}, anomalous trajectory detection~\cite{giannotti2011unveiling}, and urban region function identification~\cite{pan2012land}.
For intelligent traffic community, previous studies have addressed numerous research issues \emph{e.g.}, ~\cite{zhang2011ibat} identifies unusual driving patterns from taxi GPS traces, with applications in fraud detection and monitoring urban road networks. In~\cite{jianqin2015space}, they propose a space-time visualization method to analyze  Beijing taxi GPS data by daily operation time, driver residence, and operating patterns for understanding Beijing Taxi Operations.~\cite{chen2011real} proposes a real-time method to detect anomalous trajectories as well as identify which parts of the trajectory are responsible for its anomaly's behaviour.

\textbf{Pattern Analysis from Records of Taximeters:} The records of taximeters have been widely used to improve taxi driver's performance, \emph{i.e.}, identifying popular pickup areas~\cite{li2012prediction}~\cite{ding2016understanding}~\cite{chen2014b}~\cite{peng2012collective} and optimizing passenger-hunting routes~\cite{yuan2010t}~\cite{li2011hunting}~\cite{li2012prediction}.~\cite{zhou2017detecting} utilizes a heuristic algorithm to create maximum fraudulent trajectories from the dataset.~\cite{yazici2016modeling} examines the choices made by New York taxi drivers at JFK airport, determining pick-ups versus cruising after trips.  In~\cite{liu2013fraud} the author makes a critical observation that fraudulent taxis manipulate taximeters, inflating service distances and reported speeds. In~\cite{ou2019deep} 

\textbf{Summarizing:} To our knowledge, this paper first leverages GPS traces and records of taximeters to identify IDS activities. Therefore, we briefly review the related works about exploiting GPS traces and logs of taximeters for the other tasks.

\section{Data Pre-Processing}\label{sec:datapreprocess}

\begin{table}[t!]

  \centering
  \fontsize{5.5}{7}\selectfont
  \caption{A Pre-Processed Datum}
  \label{tab:preprocessed-data}
    \begin{tabular}{|c|c|c||c|c|c|c|}
    \hline
    \multicolumn{3}{|c||}{Taxi Trace Data}
    &\multicolumn{4}{c|}{Taxi Service Data} \cr 

    \multicolumn{3}{|c||}{(A Set of GPS Points)}
    &\multicolumn{4}{c|}{(PU/DO Points)} \cr
    \hline
    \hline
    Longitude  &Latitude   &Time Stamp      &Longitude  &Latitude & 0/1$^\dag$ &Time Stamp\cr\hline
    \end{tabular}
    \begin{tablenotes}
    \item $^\dag$ 0 means a PU point, 1 denotes a DO point.
    \end{tablenotes}
\end{table}

The dataset comprising taxi GPS and taximeter records was sourced from the Beijing Transportation Information Center\footnote{\url{http://www.btic.org.cn/xxzx/}}. Specifically, each taxi is equipped with a GPS device that transmits real-time data including longitude, latitude, time stamps, and instantaneous velocity. Concurrently, taximeter records capture service details such as time stamps for pickups (PU) and drop-offs (DO), service income, taxi occupancy status ("occupied" or "vacant"), and service distance. Both data types are relayed to a central data center via telecommunication networks.

Adherence to local taxi regulations is mandatory, requiring: 1) continuous GPS connectivity; 2) accurate GPS time stamps; and 3) taximeters in optimal condition. Non-compliance, indicated by erroneous data transmission, prompts law enforcement to notify taxi companies for corrective actions.

Data pre-processing involves two critical steps:
\begin{itemize}
\item [1.]\emph{Differentiating between two-shift and one-shift taxis:} This study focuses on identifying IDS activities in one-shift taxis, given the negligible proportion of two-shift taxis in Beijing (approximately 13 percent) and their complex behavioral patterns, such as taxi transfers and passenger pickups~\cite{zhang2014understanding}.
\item [2.]\emph{Geo locating PU/DO events by matching time stamps:} With accurate time stamps from both taximeters and GPS devices, a PU/DO event's location is determined by assigning it the GPS coordinates of the nearest time-stamped GPS point.
\end{itemize}
There is no specific data filtering method except that we remove error GPS data if the number of digits after the point is less than 6. The items of a pre-processed datum are listed in Table.~\ref{tab:preprocessed-data}.


\section{METHODOLOGY}\label{sec:methodology}

\subsection{Problem Definition}

Consider a taxi $c$ operating within an urban environment. Let $\phi_T$ and $\phi_R$ represent the feature extraction functions from the GPS trace dataset and taximeter record dataset, respectively. The process of identifying taxis engaged in the IDS activities is formalized as follows:

\begin{mydef}[Identification of IDS Activities]
Given feature extractors $\phi_T$ and $\phi_R$, the objective is to determine a function $f$ that classifies a taxi's involvement in IDS activities, i.e.,
\begin{equation}\label{eqt:identify-SDA}
f\left(\phi_T(c) ,\phi_R(c)\right) \rightarrow {+1,-1},
\end{equation}
where $+1$ signifies the presence and $-1$ the absence of IDS activities in a taxi.
\end{mydef}

Equation~\ref{eqt:identify-SDA} frames the identification of IDS activities as a supervised learning challenge. This study distinguishes between positive and negative samples for classification, leveraging Beijing's "taxi role model" initiative to define role model taxis as negative samples $\mathbf{\Omega}-$, and taxis exhibiting IDS activities as positive samples $\mathbf{\Omega}+$.

\textbf{IDS Discovery Framework:} The proposed methodology encompasses three main stages:
\begin{itemize}
\item [] \textbf{Step 1. Heterogeneous Driver Behavior Modeling:} The STL (Sleeping Time and Location) and PU (Pick-Up) behaviors are computed to encapsulate individual driver behaviors.
\item [] \textbf{Step 2. Multi-Scale Pooling from Self-Similarity:} Individual behaviors are translated into a unified feature space via Self-Similarity (SS) and Multi-Scale Pooling (MSP).
\item [] \textbf{Step 3. Supervised Learning for IDS Activity Identification:} Leveraging the insights from the initial stages, the IDS detection problem is approached through supervised classification techniques.
\end{itemize}

\section{Heterogeneous Driver Behavior Modeling}\label{sec:behavior-model}

\subsection{Modeling Sleeping Behavior of a Driver}
\begin{figure}[t!]
  \centering
  \subfigure[A Taxi with the plate ``9689'']{
    \label{fig:subfig:one-taxi} 
    \includegraphics[width=.23\textwidth]{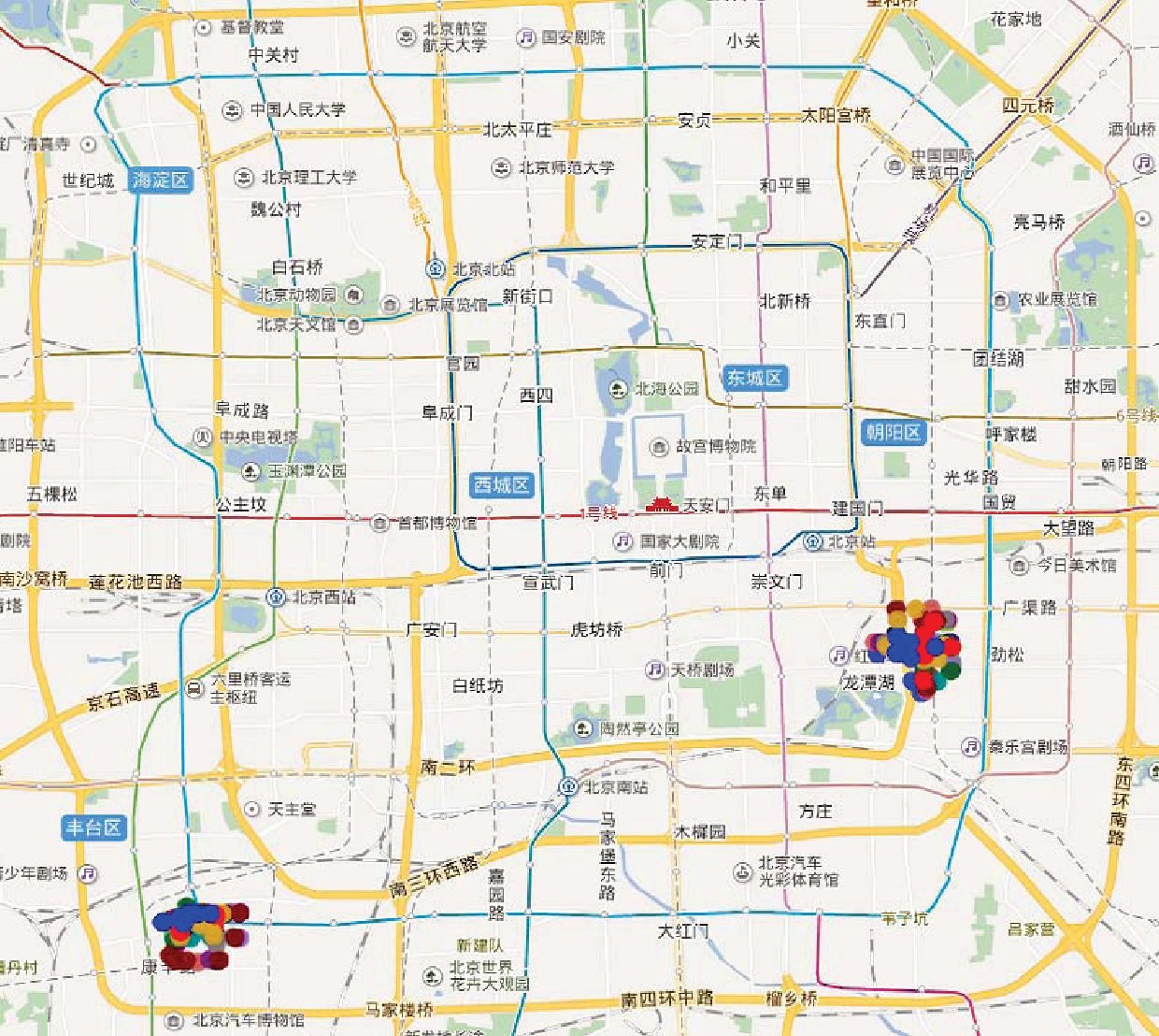}}
  \subfigure[A Taxi with the plate ``8294'']{
    \label{fig:subfig:two-taxi} 
    \includegraphics[width=.223\textwidth]{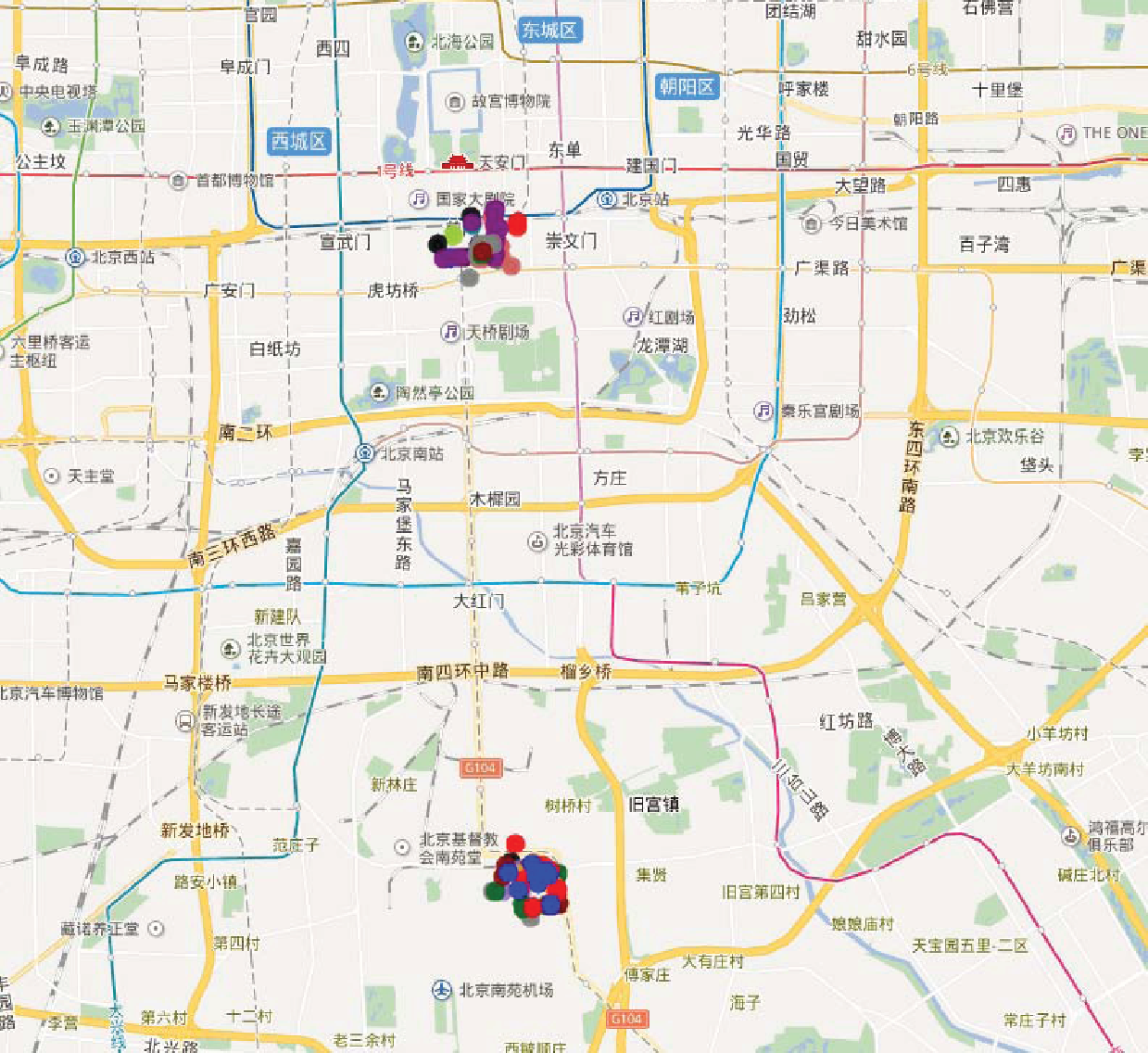}}
  \caption{An illustration of the sleeping locations of taxis drivers.}
  \label{fig:taxi-STL} 
\end{figure}

\begin{mydef}[Definition of STL]\label{def:sleepingtimeandlocation}
STL for a taxi driver is characterized by a period of inactivity, where the taxi's location remains unchanged, and it is unoccupied. This period, denoted as the driver's sleep, is defined by three parameters: the location of rest, the starting time of this inactive period, and its duration. Hence, STL is formally represented as a triple: \emph{sleeping location, start time, and sleep duration}.
\end{mydef}

STL can be straightforwardly derived for each taxi based on Definition~\ref{def:sleepingtimeandlocation}. Fig~\ref{fig:taxi-STL} depicts the sleep locations of two one-shift taxis in Beijing, highlighting the distinct, concentrated sleeping patterns that reflect personal domiciles. Conversely, the sleeping behaviors of two-shift taxi drivers are more varied, including breaks at random locations and rest within the taxi, making STL modeling less applicable.

Given a taxi's start time $t_s$, sleeping coordinates $(p_{lon},p_{lat})$, and sleeping duration $t_d$, STL is vectorized as $\mathbf{x}^{STL}=[t_s, p_{lon},p_{lat}, t_d]^\top$. For efficient and accurate classification, the Fisher Vector (FV) method is employed to transform STLs into discriminative features, a technique proven effective in image recognition tasks~\cite{karpathy2014large}.

\textbf{Encoding STL by Fisher Vector.}
In a nutshell, FV assumes that the gradient of the log-likelihood describes the contribution of the parameters $\lambda$ with respect to the generation of data $X$ in a parameter space. Concretely, FV firstly fits a Gaussian Mixture Model (GMM). The resulting FV descriptor integrates the deviations of the parameters from GMM, providing a robust feature~\cite{song2017low}. FV's efficacy spans a range of imaging applications, including image classification, face recognition, object detection, and texture analysis~\cite{jain201515}\cite{simonyan2013fisher}\cite{chatfield2014return}~\cite{cimpoi2014describing}. 


Let $X =\{\mathbf{x}^{STL}_i,i = 1 \ldots N\} $ be a set of STLs from $N$ taxis. FV $G^X_\lambda$ is modeled as the gradients of a probability density function $u_\lambda$ as follows:
\begin{equation}\label{eqt:fv}
G^X_\lambda = \frac{1}{T}\nabla_\lambda\log u_\lambda(X).
\end{equation}
where $T$ is the number of STL data in a time bucket since some one-shift taxis occasionally have no sleep time. In practice, $T$ is almost constant in our experiment.

In practice, following~\cite{sanchez2013image}, we choose GMM to be $u_\lambda$ which approximates with arbitrary precision to any continuous distributions: $u_{\lambda}(\mathbf{x})=\sum_{k=1}^K w_kp_k(\mathbf{x})$ with $w_k\geq 0, \sum_{k=1}^Kw_k=1$, in which the parameters $\lambda =\{\bm{\mu}_k,\bm{\Sigma}_k\}, k=1,\ldots, K$, where $\bm{\mu}_k$ and $\bm{\Sigma}_k$ are respectively the mixture weight, the mean vector and the covariance matrix of the $k$-th Gaussian component $p_k(\mathbf{x})$:
\begin{small}
\begin{equation}
p_k(\mathbf{x})=\frac{1}{(2\pi)^{D/2}{|\bm{\Sigma}_k|}^{1/2}}\text{exp}\left\{-\frac{1}{2}(\mathbf{x}-\bm{\mu}_k)^\top\bm{\Sigma}^{-1}_k(\mathbf{x}-\bm{\mu}_k)\right\},
\end{equation}
\end{small}

We assume that the covariance matrix $\bm{\Sigma}_k$ is diagonal matrix since the computational cost of the diagonal covariances is much lower than the cost involved by full covariances. Hereafter, we use the notation $ \sigma_k^d $ represent the $d$-th elements on the diagonal in the $k$-th covariance matrix.
For the weight parameters $w_k$, we adopt the soft-max formalism in~\cite{sanchez2013image} and define $w_k=\frac{\exp(\alpha_k)}{\sum_{i=1}^{K}\exp(\alpha_k)}$, where the re-parametrization using the $\alpha_k$ avoids enforcing explicitly the constraints in GMM. Consequently, the gradient of log-likelihood $\mathcal{L}(X|\lambda)$ with respect to the parameters $\mu_k^d$ ($\bm{\mu}_k=[\mu_k^1,\ldots,\mu^d_k,\ldots, \mu_k^D]^\top$) and $\sigma_k^d$ ($\bm{\Sigma}_k= diag([\sigma^1_k,\ldots, \sigma_k^d,\ldots, \sigma_k^D]^T)$) is respectively as follows:
\begin{equation}
\frac{\partial \mathcal{L}(\mathbf{x}_i^{STL}|\lambda)}{\partial\mu_k^d} = \gamma_i(k) \left( \frac{\mathbf{x}_i^{STL^d} - \mu_i^d}{(\sigma_k^d)^2} \right)\label{eqt:FV-lambda},
\end{equation}
\begin{equation}
\frac{\partial \mathcal{L}(\mathbf{x}_i^{STL}|\lambda)}{\partial\sigma_k^d} =  \gamma_i(k)\left( \frac{(\mathbf{x}_i^{STL^d} - \mu_k^d)^2}{(\sigma_k^d)^3} -\frac{1}{\sigma_k^d}\right)\label{eqt:FV-sigma},
\end{equation}
where $\gamma_i(k)$ is denoted as the probability of a STL $\mathbf{x}_i^{STL}$ point to be generated from the $k$-th Gaussian component,
\begin{equation}
\gamma_i(k)=\frac{w_kp_k(\mathbf{x}^{STL}_i|\lambda)}{\sum_{k=1}^Kw_kp_k(\mathbf{x}_i|\lambda)}.
\end{equation}
Consequently, following the principle in~\eqref{eqt:fv}, the FV of a $\mathbf{x}$ is the concatenation of the partial derivatives with respect to the mean $\mu_k^d$ and the standard deviation $\sigma_k^d$ as follows:
\begin{small}
\begin{equation}\label{eqt:fishervector}
\mathbf{f}^{STL}_i=\frac{1}{T}\left[\frac{\partial \mathcal{L}(\mathbf{x}|\lambda)}{\partial\mu_k^1},\ldots,\frac{\partial \mathcal{L}(\mathbf{x}|\lambda)}{\partial\mu_k^D}, \frac{\partial \mathcal{L}(\mathbf{x}|\lambda)}{\partial\sigma_k^1},\ldots,
\frac{\partial \mathcal{L}(\mathbf{x}|\lambda)}{\partial\sigma_k^D}\right]^\top
\end{equation}
\end{small}
where $D$ is the dimension of the STL vectors. By~\eqref{eqt:fishervector}, a STL point $\mathbf{x}^{STL}$ is encoded into a FV vector with $2DK$ dimension. Following the  $\ell_2$-normalization~\cite{karpathy2014large}, we finally use the normalized FV.

\subsection{Modeling the PU Behavior of a Driver}

\textbf{Developing the Latent Pickup (PU) Behavior Model:} It's well-documented that individuals exhibit consistent spatio-temporal patterns, a concept that extends to the predictable behaviors of taxi drivers regarding passenger pickups~\cite{wang2017human}. The underlying principle is that drivers quest for profitability leads to the emergence of distinct pickup patterns. Notably, drivers with the IDS activities tend to operate within specific zones known for high ride service demand. This behavior contrasts with the registered drivers, whose pickup locations are generally more dispersed and appear random~\cite{fazio1981direct}~\cite{phiboonbanakit2016does}.

We represent a pickup event with $\mathbf{x}^{PU}=[t_{pu},p_{lon},p_{lat}]^\top$, capturing the essential details of when and where passengers are picked up. This study employs LDA to model the latent structures within the pickup data, drawing parallels between the distribution of words in documents and pickup points in urban spaces:
\begin{itemize}
\item The aggregation of pickup points resembles the collection of words within a document.
\item The compilation of pickups over a specific period forms our corpus, analogous to the textual content of a document.
\item The entirety of pickups by a single taxi is viewed as an individual document in this analogy.
\end{itemize}
This framework allows us to employ latent topics to elucidate the patterns in drivers pickup behaviors, with the specific analogy detailed in Table~\ref{tbl:correspondenceinLDA}. LDA leverages a dirichlet prior to model the distribution of topics within documents, illustrating its versatility in representing texts. Documents can thus embody a mix of multiple topics, enhancing the model's descriptive power~\cite{wang2007spatial}~\cite{anandkumar2012spectral}. 

Consider $\mathcal{D}$, a collection of $M$ such "documents" $\mathcal{D}={\mathbf{w}_1,\mathbf{w}_2,\ldots, \mathbf{w}_M}$, each "document" $\mathbf{w}$ being a series of "words" $\mathbf{w}={w_1,w_2,\ldots,w_N}$, or in another notation, $w_{1:N}$. These "documents" are presumed to arise from a distribution $\bm{\theta} =[\theta_1,\ldots,\theta_T]^\top$ over a set of topics.

To delineate the "words" representing PU behavior, this work utilizes the $k$-means algorithm to cluster the pickup points $\mathbf{x}^{PU}_i$ from all taxis into a designated number of clusters. Each cluster represents a "word" $w_n$, which, when aggregated daily, forms a "document" that characterizes the pickup behavior of a driver.


\begin{table}[t!]
\centering
\small{
\caption{The correspondence between the text corpus and the PU behavior.}~\label{tbl:correspondenceinLDA}
\begin{tabular}{|c|p{3cm}|p{3cm}|}\hline
Notation & Text Corpus & PU Points \\[0.5ex]\hline\hline
$\mathbf{z}$    & Topics        &  The PU Behavior\\
$\mathbf{w}$ & A document & A Time bucket  \\
$\theta$ & Topic proportions & Behaviors Proportions \\
\hline
\end{tabular}
}
\end{table}

The joint distribution of a topic mixture $\bm{\theta}$, a set of topics $\mathbf{z}$, and a set of $N$ words $\mathbf{w}$ is given as follows:
\begin{equation}
p(\bm{\theta},\mathbf{z},\mathbf{w}|\alpha, \beta)=p(\bm{\theta}|\alpha)\prod_{n=1}^N p(z_n|\bm{\theta})p(w_n|z_n,\beta),
\end{equation}
where $p(\bm{\theta}|\alpha)$ follows a Dirichlet distribution with the simplex parameter $\alpha$ (\emph{i.e.}, $\alpha$ makes $\theta_i\geq0,\sum_{i=1}^T\theta_i=1$), $p(z_n|\bm{\theta})$ is simply $\theta_i$ for the unique $i$ such that $z_n^i=1$, and $p(w_n|z_n,\beta)$  is a multinomial probability conditioned on the topic $z_n$.

We aim to determine the latent topics based on the PU points in a time bucket. This is equivalent to computing the posterior distribution of the hidden variables given a document:
\begin{equation}
p(\bm{\theta},\mathbf{z}|\mathbf{w},\alpha,\beta)=\frac{p(\bm{\theta},\mathbf{z},\mathbf{w}|\alpha,\beta)}{p(\mathbf{w}|\alpha,\beta)},
\end{equation}
which is approximately computed based on the Monte Carlo Markov Chain (MCMC) technique~\cite{blei2003latent} where the idea is to generate posterior samples from its conditional distribution. Therefore, the latent topics of the PU behavior $\mathbf{f}^{PU}$ for a taxi in a day is defined as follows:
\begin{equation}\label{eqt:pickupfeature}
\mathbf{f}^{PU}=[p(\theta_1,\mathbf{z}|\mathbf{w},\alpha,\beta),\ldots,p(\theta_k,\mathbf{z}|\mathbf{w},\alpha,\beta)]^\top
\end{equation}



\textbf{Dealing with location inaccuracy:}  In many cases, the location information may need to be more accurate due to the blocked signal transmission caused by high buildings or overpasses. Therefore, a smooth mechanism is preferred over the \emph{degree of PU word membership} should be performed during computing the words $\mathbf{w}$ for a taxi.

A simple and yet reasonable way is to use the Nadaraya-Watson kernel regression on the Gaussian kernel to approximate the probabilistic word membership~\cite{shapiai2010non}, which we denote as $o$. The probabilistic word of a PU point is as follows:
\begin{equation}
\begin{split}
&o_a= \frac{\exp(-d^2_a/\delta)}{\sum \exp(-d^2_i/\delta)}, \  \ o_b = \frac{\exp(-d^2_b/\delta)}{\sum \exp(-d^2_i/\delta)}\\
& o_a+o_b =1.
\end{split}
\end{equation}
where $d$ is the distance between a PU point and a center of a PU word, and $\delta$ denotes the uncertain range of a sensor.

\subsection{Multi-Scale Pooling for Self-Similarity Based Features}

\begin{figure}[t!]
  \centering
  \subfigure[Discover and align the changes of behaviors by SS and pooling]{
    \label{fig:subfig:SS} 
    \includegraphics[width=.4\textwidth]{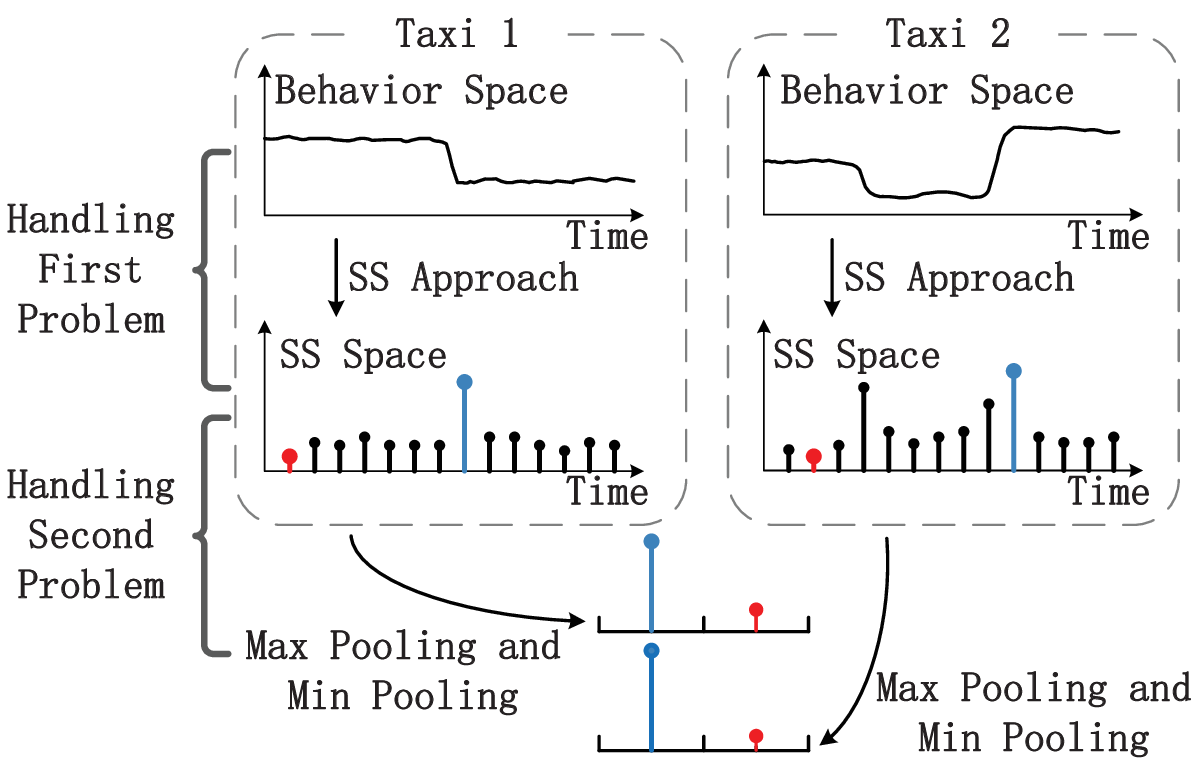}}
  \subfigure[Assemble aligned feature vectors into a discriminative feature]{
    \label{fig:subfig:msp-pooling} 
    \includegraphics[width=.48\textwidth]{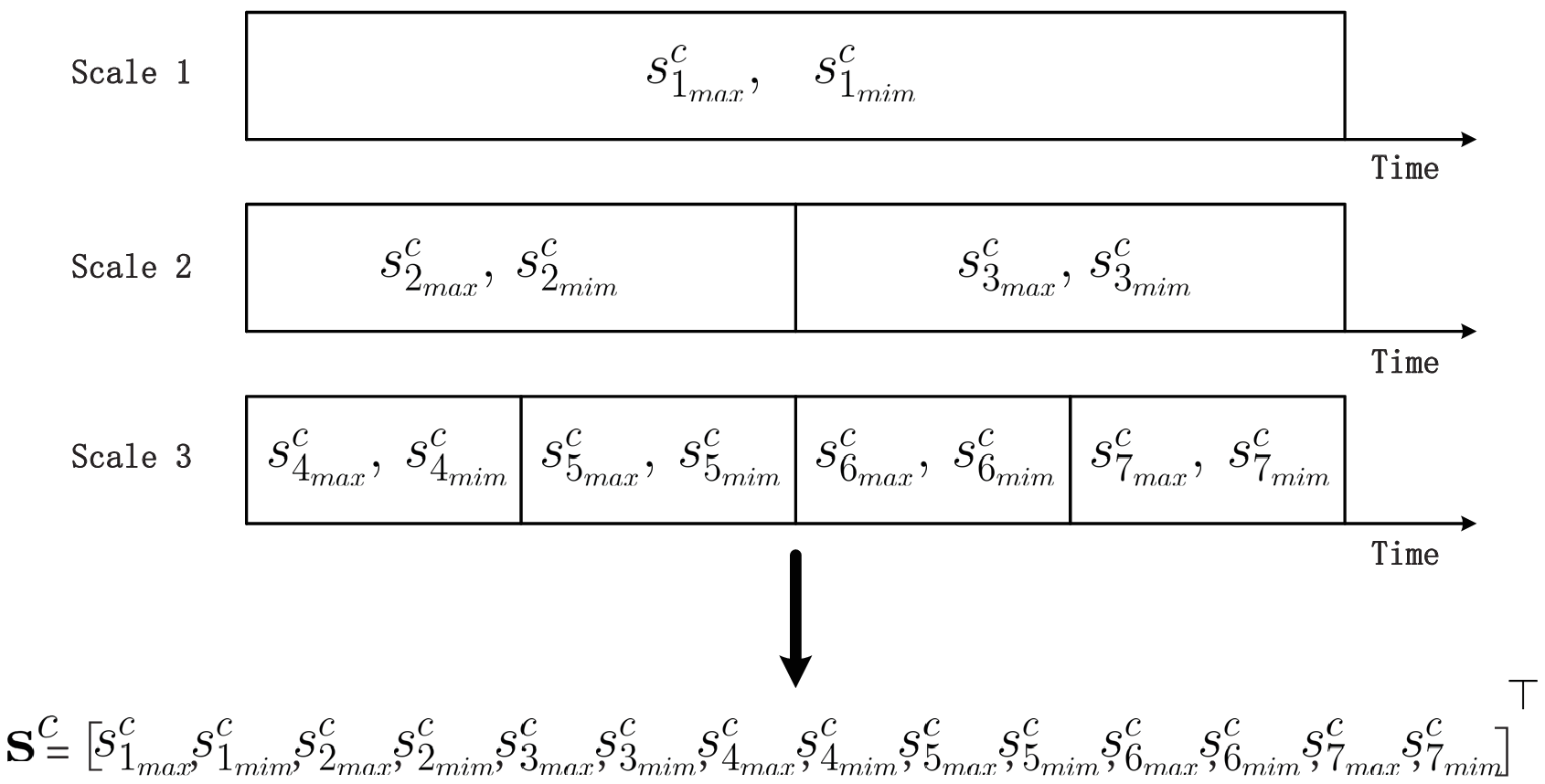}}
  \caption{An illustration of MSP pooling for SS-based features.}
  \label{fig:msp-pooling-feature} 
\end{figure}

It is posited that individual taxi drivers exhibit distinct, yet relatively stable, behavioral patterns. Anomalies in these patterns, especially sharp deviations observed over short intervals (e.g., daily), may signal the IDS activities. Identifying such anomalies necessitates addressing two primary challenges:

\begin{itemize}
\item \emph{Detecting Significant Behavioral Changes:} Variability in a driver's behavior can naturally occur due to unforeseen events. It is crucial, therefore, to establish a baseline against which significant deviations indicative of potential IDS activities can be measured. 
\item \emph{Consolidating IDS Indicators Across Taxis:} IDS activities, when present in multiple taxis, may not manifest simultaneously. As depicted in Fig.~\ref{fig:subfig:SS}, aligning these activities within a unified feature vector framework is essential. 
\end{itemize}

For the first problem, Self-Similarity (SS) approach is proposed to detect the occurrence of the IDS activity. Concretely, given a sequential state of behaviors $\mathbf{f}^c_{b_i}$ (which are either the STL behavior $ \mathbf{f}^{STL}_{c_{b_i}}$ or the PU one $\mathbf{f}^{PU}_{c_{b_i}}$) of a taxi $c$ at the $i$-th day in a time bucket $b$, SS computes the difference of the behaviors between two consecutive days as follows:
\begin{equation}\label{eqt:self-similarity}
s^c_{b_{i,j}} = Similarity\left(\mathbf{f}^c_{b_i},\mathbf{f}^c_{b_j}\right).
\end{equation}
There are two important parameters in~\eqref{eqt:self-similarity}: the function $Similarity(\cdot,\cdot)$ and the time bucket $b$. The function $Similarity(\cdot,\cdot)$ can be \emph{any} function that measures the similarity between two features. For instance, mutual information is used for two distributions, and cosine distance is for two normalized vectors. The time bucket $b$ in~\eqref{eqt:self-similarity} corresponds to the time window in which the IDS activity maybe occur. Ideally, if the length of the time bucket is larger, the IDS activity is much easier to cover. This paper chooses cosine distance as the similarity function, and the bucket length is 30 days.

To the second problem, taking the ``Taxi 1'' in Fig.~\ref{fig:subfig:SS} as an example, if the change of the driver behaviors is detected by the SS approach, the maximal change could be used to indicate the occurrence of the IDS activity. Therefore, given a time bucket $b$, the occurrences of the IDS activity are discovered by max pooling as follows:
\begin{equation}\label{eqt:maxpooling}
s^c_{b_{max}} = \text{max}\{s^c_{b_{i,j}}\}.
\end{equation}

Meanwhile, the driver behaviors, especially the PU behaviors, are only sometimes consistent due to some unexpected events. For instance, picking up a passenger via the car-hailing service or changing a domicile because the rent is due. To describe this observation, the minimal change is extracted to calibrate the maximal change as follows:
\begin{equation}\label{eqt:minpooling}
s^c_{b_{min}} = \text{min}\{s^c_{b_{i,j}}\}.
\end{equation}
The combination of~\eqref{eqt:maxpooling} and~\eqref{eqt:minpooling} jointly describes the intense change of the behaviors.

As illustrated in Fig.~\ref{fig:subfig:SS}, although the changes of the driver behaviors occur at different days, both max pooling~\eqref{eqt:maxpooling} and min pooling~\eqref{eqt:minpooling} encode the IDS activity into the same bin of the feature vector. Ideally, by combining the SS approach and the pooling method, the individual-wise behaviors are aligned into a universal feature space for all taxies.

When the length of the time bucket is very large, some unexpected events tend to occur. Both max pooling and min pooling would capture the unexpected events rather than the IDS activity. Therefore, Multiple time-Scale Pooling (MSP) is further proposed to increase the discriminative power of the aligned features. Concretely, Fig.~\ref{fig:subfig:msp-pooling} shows that a time bucket is divided into multiple smaller ones by several scales, and then the aligned feature vectors from different scales are concatenated into a MSP-SS vector as follows:
\begin{equation}\label{eqt:ss-based-feature}
\mathbf{s}^c=\left[s^c_{1_{max}},\ldots,s^c_{B_{max}},s^c_{1_{min}},\ldots,s^c_{B_{min}}\right]^\top,
\end{equation}
where $B$ is the number of the divided time buckets.

Intuitively, when an unexpected event occurs at the $6$-th time bucket, the top scale, \emph{i.e.}, scale 1 in Fig.~\ref{fig:subfig:msp-pooling}, may consider the unexpected event as the IDS activity. Because the pooling operations in~\eqref{eqt:maxpooling} and~\eqref{eqt:minpooling} discard the structure information. In a contrast, if we introduce more scales (\emph{e.g.}, the scale 2 and the scale 3 in Fig.~\ref{fig:subfig:msp-pooling}), the other time buckets (\emph{e.g.}, the $2$-th) still capture the occurrence of the IDS activity. As a result, MSP increases the the discriminative power of the feature.

\subsection{Multiple Component Learning for Aligning IDS Behavior}

Two problems are raised to classify the taxis with IDS behavior as follows:
 \begin{itemize}
 \item How to align and classify the behavior in the long-range time bucket is key to find the taxis with IDS, although MSP could align the behavior in a short-range time bucket(\emph{e.g.}, a week).
 \item  If one of the STL-based feature, the PU-based one and the combination of the two behaviors is deficient, how to identify the taxi with IDS activity?
 \end{itemize}
This paper proposes a hybrid approach to align and leverage features in a long-range time bucket by combining Multiple Instance Learning (MIL) and Multiple Component Learning (MCL)~\cite{kim2008mcboost}.

Let the behaviors of a taxi in a long-range time bucket $T$ be represented as a super feature set $\mathcal{X}_i=\left\{\mathcal{S}_i^k \right\}_{k=1}^K$
, where $k$ is the number of the behaviors. In this paper, the feature set $\mathcal{S}_i^1=\{\mathbf{s}^{STL}_{it}\}^{T_{STL}}_{t=1}$ and the set $\mathcal{S}_i^2=\{\mathbf{s}_{it}^{PU}\}^{T_{PU}}_{t=1}$, in which
$\mathbf{s}^{STL}_{it}$ and $\mathbf{s}^{PU}_{it}$ is the $t$-th time bucket of time range $T_{STL}$ and $T_{PU}$, respectively. The number of time bucket $T_{STL}$ and $T_{PU}$ respectively are not necessary equal to $T$, \emph{i.e.}, $T_{STL} \leq T$, and $T_{PU} \leq T$. The proposed Multiple Component-Multiple Instance Learning (MC-MIL) is as follows:
\begin{equation}\label{eqt:MCL}
P_i=1-\prod_{k=1}^K(1-P^k(\mathcal{X}_i))
\end{equation}
where $k$ is the number of classifier, $1\leq k\leq K$, and function $P^k(\cdot)$ outputs the probability of a taxi with the IDS activity. In our work, $K=2$ represents both the classifiers built from the STL feature and the PU one, respectively.~\eqref{eqt:MCL} is the result-fused approach which combines the results of $K$ classifiers $P^k(\cdot)$ into $P_i$. The advantage of~\eqref{eqt:MCL} is that if any one of the behavior feature is missed,~\eqref{eqt:MCL} still robust predicts a result.

The $P^k(\cdot)$ is MIL function to align the IDS behavior as follows:
\begin{equation}\label{eqt:MIL}
P^k(\mathcal{X}_i)=1-\prod^T_{t}(1-P^k_{t}(\mathcal{S}_{i}^k))
\end{equation}
where the $t$ is the index of a time bucket to compute MSP~\eqref{eqt:ss-based-feature}.~\eqref{eqt:MIL} utilizes the noise-or model to align the IDS behavior among a long-range bucket $T_{STL}$ and $T_{PU}$, respectively. The probability $P^k_{t}(\cdot)$ is logistic regression as follows:
\begin{equation}
P^k_{t}(\mathcal{S}^k_i)= \frac{1}{1+\exp(-H^k(\mathbf{s}^k_{it}))}
\end{equation}
where $H^k_i(\mathbf{s}^k_{it})$ is the additive function as follows:
\begin{equation}
H^k(\mathbf{s}^k_{it})=\sum_{j=1}^N \alpha^k_{j} h^k_j(\mathbf{s}^k_{it})
\end{equation}
where $\mathbf{s}^k_{it}$ is the $k$ SS-based behavior feature in the $t$ time bucket for the $i$-th taxi, $h^k_j(\mathbf{s}^k_{it})$ is a weak classifier, $h^k_j(\mathbf{s}^k_{it})\in \{-1,+1\}$, and $\alpha^k_{j}$ is the non-negative coefficient, $0\leq \alpha^k_{j}$.

Under this model the likelihood assigned to a set of training bags $\mathcal{X}_i$ and its label $y_i, y_i\in \{0,1\}$ is as follows:
\begin{equation}
L(H^k_j(\mathbf{s}^k_{it}))=\prod_{i=1}^N P_i^{y_i}(1-P_i)^{(1-y_i)}
\end{equation}
Following AnyBoost approach~\cite{ll1999boosting}, the weight on each instance $\mathbf{s}^k_{it}$ is given as the derivative of the cost function with respect to a change in the score of the example. The derivative of the log likelihood~\footnote{For clarity, we abbreviate the functions $H^k(\mathbf{s}^k_i)$ as $H^k$, $P^k_{t}(\mathcal{S}^k_i)$ as $P^k_{t}$, and $P_k(\mathcal{X}_i)$ as $P_k$.} is:
\begin{equation}\label{eqt:MC-MIL-weight}
\frac{\partial L(H^k)}{\partial H^k_i} =
\frac{\partial L(H^k_i)}{\partial P_i}\cdot
\frac{\partial P_i}{\partial P_k}\cdot\frac{\partial P_k}{\partial P^k_t}\cdot\frac{\partial P^k_t}{\partial H_i^k}
\end{equation}
where the derivations are computed as follows:
\begin{eqnarray}
\frac{\partial L(H^k)}{\partial P_i} &=& \frac{y_i}{P_i}+\frac{y_i-1}{1-P_i} \\
\frac{\partial P_i}{\partial P_k} &=& \prod_{j\neq k}^K(1-P^j)\\
\frac{\partial P_k}{\partial P^k_t} &=&\prod_{j\neq t}^T(1-P_j^k) \\
\frac{\partial P^k_t}{\partial H^k} &=& P^k_t(1-P^k_t)
\end{eqnarray}
Therefore,~\eqref{eqt:MC-MIL-weight} is simplified as follows:
\begin{equation}
\frac{\partial L(H^k)}{\partial H^k_i}= \frac{y_i-P_i}{P_i}P^k_t
\end{equation}
The parameter $\alpha^k_{j}$ is determined using a line search to maximize $\log L(H^k+\alpha^k_{j})$. During the implementation, Classification And Regression Trees (CARTs)~\cite{cart-breiman-84} are used to built the weak classifier $h^k_j(\mathbf{s}^k_{it})$.

\begin{table*}[t!]
  \centering
  \caption{Effectiveness of PU-based SS features when the number of Words is 200.}
  \label{tab:effectiveness-PU-when-word-fixed}
  \begin{tabular}{|c||c|c||c|c||c|c||c|c||c|c||c|c|}
    \hline
    \multirow{2}{*}{\# Topics} & \multicolumn{2}{c||}{LSTM} & \multicolumn{2}{c||}{Transformer} & \multicolumn{2}{c||}{RF} & \multicolumn{2}{c||}{GBT} & \multicolumn{2}{c||}{MKL} & \multicolumn{2}{c|}{LR} \\ \cline{2-13}
     & AUC & AP & AUC & AP & AUC & AP & AUC & AP & AUC & AP & AUC & AP \\
    \hline
    20 & 0.8647 & \textbf{0.4851} & 0.8413 & 0.6014 & 0.9080 & 0.6947 & 0.8929 & 0.6355 & \textbf{0.9024} & 0.5860 & 0.5726 & 0.2061 \\
    40 & 0.8253 & 0.4526 & 0.8517 & 0.6148 & 0.8783 & 0.6701 & 0.8919 & 0.5865 & 0.8003 & 0.5738 & 0.5445 & 0.3470 \\
    60 & \textbf{0.8780} & 0.4719 & 0.8319 & \textbf{0.6254} & 0.9196 & \textbf{0.7676} & 0.9095 & \textbf{0.8010} & 0.8364 & 0.6038 & 0.6647 & 0.4706 \\
    80 & 0.8426 & 0.4613 & 0.8501 & 0.6024 & 0.9085 & 0.7338 & \textbf{0.9206} & 0.7685 & 0.8571 & \textbf{0.6447} & \textbf{0.7778} & \textbf{0.5696} \\
    100 & 0.8516 & 0.4698 & \textbf{0.8629} & 0.6028 & \textbf{0.9306} & 0.6701 & 0.8994 & 0.7044 & 0.8260 & 0.5837 & 0.6687 & 0.2497 \\
    \hline
  \end{tabular}
\end{table*}

\section{Experiments and Discussion}\label{sec:expeirments}

\subsection{Experiment Setup}

\textbf{Real Data.} We release dataset data set for our experiments publicly and any one could access it here.\footnote{https://github.com/pangjunbiao/IDS-BJ}. The data set, referred as ``IDS@BJ'', includes the initial information for each taxi individually such as (plate Id, origin time, longitude, latitude, Destination time, destination longitude, destination latitude). The time of identifying one-shift taxis with the IDS activity ranges from Jan. 2015 to Seq. 2016. Because the taxis with IDS activity are sparse, manually determining whether a randomly picked taxi has IDS activity depends on a law enforcer's personal experience.

\textbf{Evaluation metrics:} In our experiments, we use two kinds of evaluation criteria to evaluate the effectiveness as follows:
\begin{itemize}
\item [1.]\emph{Precision and Recall Curve (PRC):} PRC plots Precision versus Recall of a classifier at different decision thresholds. 
The Averaged Precision (AP) of the PRC is used to numerically indicate the performance of a detection system. The higher AP of PRC is, the better a classifier is~\cite{prc-web}.
\item [2.]\emph{Receiver Operating Characteristic (ROC) curve:} ROC curve plots True Positive Rate (TPR) versus False Positive Rate (FPR) of a classifier at different decision thresholds. 
Area Under the Curve (AUC) is further used to evaluate the performance of a classifier. The higher AUC is, the better a classifier is. 
\end{itemize}
Due to the imbalance of the test set, AP is more reasonable than AUC to evaluate the performance of classifiers in this paper.

\textbf{Experiment Settings.} Both the STL behavior and the PU behavior are extracted from each day. Three scales of MSP are built as in Fig.~\ref{fig:subfig:msp-pooling}, \emph{i.e.}, $1\times 16$ days, $2\times 8$ days, and $4\times 4$ days. The SS function in~\eqref{eqt:self-similarity} is defined as cosine distance between two features. To better align the IDS behavior, the long rang time bucket $T=30$, a $16$ days sliding widow with step 4 is used to split the $T=30$ into 26 overlapped SS features. That is, $T_{STL}$ and $T_{PU}$ are equal to 26, respectively.

The IDS discovering system is implemented under Python 2.7 on a 3.30 GHz machine with 8G RAM.

\subsection{Effectiveness on the SS-based Behavior Features}

\textbf{Baseline Classifiers.} We evaluate our behavior-based features on the following baseline classifiers:
\begin{itemize}
\item [1.] \textbf{Long Short-Term Memory (LSTM)~\cite{staudemeyer2019understanding}:} LSTM is a type of Recurrent Neural Network (RNN) capable of learning long-term dependencies for sequential data. The LSTM model was configured with the following hyper-parameters: learning rate is $1 \times 10^{-3}$, weight decay for regularization is $1 \times 10^{-5}$, batch size is 64, and the Adam optimizer with a reduce learning rate on plateau scheduler reduces the learning rate by a factor of $0.5$ with a patience of 5 epochs.
\item [2.] \textbf{Transformer~\cite{liu2021transformer}:} Transformer utilizes self-attention mechanisms to obtain a highly effectiveness for classification tasks which involve complex relationships within the data~\cite{han2021transformer}. We utilized Adam optimizer with a weight decay of $1 \times 10^{-5}$ for regularization. The learning rate scheduler was reduce learning rate on plateau, with a reduction factor of 0.5 and a patience of 5. The batch size is 32.
\item [3.] \textbf{Logistic  Regression (LR):} LR is a widely used statistical model that uses a logistic function to model a binary dependent variable.
\item [4.] \textbf{Random Forest (RF)~\cite{breiman2001random}:} RF, an ensemble of CARTs~\cite{cart-breiman-84}, is trained with the bagging method.  Compared with CARTs, RF tends to prevent the over-fitting by averaging a set of ``shallow'' CARTs.
There are four important hyper-parameters, \emph{i.e.}, maximum tree depth, maximum number of leaf nodes, minimum number of points per leaf node, and number of trees, in RF. These hyper-parameters are chosen by random searching method~\cite{bergstra2012random}\footnote{During the random searching, maximum tree depth ranges from 2 to 4; maximum number of leaf nodes ranges from 2 to 8; minimum number of points per leaf node ranges from 1 to 3; number of trees ranges from 50 to 100.}. Meanwhile, the number of features to consider per split in a decision tree is the square root of the total number of features~\cite{nan2015feature}.
\item [5.] \textbf{Gradient Boosting Trees (GBT)~\cite{friedman2001greedy}:} GBT combines a set of weak CARTs into a single strong learner in an iterative fashion. Compared with RF and CARTs, GBT allows arbitrary differentiable loss functions to be used. In this paper, the exponential loss $e^{-yf(\mathbf{s})}$ is used due to the excellent binary classification ability.
Specially, the tree-specific parameters are determined by the random searching~\cite{bergstra2012random}, while the boosting parameters are selected by 5-fold cross validation\footnote{During the cross validation, the learning rate ranges from 0.1 to 0.5; and the number of sequential trees ranges from 50 to 100.}.
\item [6.] \textbf{Multiple Kernel Learning (MKL)~\cite{gonen2011multiple}:} MKL linearly combines multiple kernels into Support Vector Machines (SVMs), efficiently fusing multiple types of features into a kernel; besides, MKL can automatically determine which kernels are useful. In this paper, MKL uses 5 different types of kernels, \emph{i.e.}, polynomial kernel, Gaussian kernel, linear kernel, intersection kernel, and Chi-squared kernel. There are 10 ($5\times 2$) kernel matrices which are used in MKL. 
\end{itemize}

We aim to demonstrate that the proposed features are discriminative in identifying the nuanced IDS activities across a range of classifiers from the classical methods to the advanced neural network architectures.

\begin{table*}[t!]
  \centering
  \caption{Effectiveness of STL-based SS features on different baseline classifiers.}
  \label{tab:effectiveness-STL}
  \begin{tabular}{|c||c|c||c|c||c|c||c|c||c|c||c|c|}
    \hline
    \multirow{2}{*}{\# Gaussian} & \multicolumn{2}{c||}{LSTM} & \multicolumn{2}{c||}{Transformer} & \multicolumn{2}{c||}{RF} & \multicolumn{2}{c||}{GBT} & \multicolumn{2}{c||}{MKL} & \multicolumn{2}{c|}{LR} \\ \cline{2-13}
    & AUC & AP & AUC & AP & AUC & AP & AUC & AP & AUC & AP & AUC & AP \\
    \hline
    2 & 0.7607 & 0.4324  & 0.8141 & 0.3119 & \textbf{0.8248} & 0.4008 & 0.7468 & 0.3028 & 0.7290 & 0.3095 & \textbf{0.8056} & \textbf{0.4867} \\
    4 & 0.8357 & \textbf{0.4843} & 0.8417 & \textbf{0.3450} & 0.7401 & 0.4183 & 0.7495 & 0.3508 & 0.7416 & 0.4155 & 0.7356 & 0.4067 \\
    6 & 0.7972 & 0.4435 & 0.8594 & 0.3295 & 0.7182 & 0.3707 & 0.6875 & 0.3790 & 0.7553 & 0.3570 & 0.7014 & 0.3376 \\
    8 & \textbf{0.8795} & 0.4600 &\textbf{0.8674} & 0.3197 & 0.7607 & \textbf{0.5316} & \textbf{0.7698} & \textbf{0.4880} & \textbf{0.7736} & 0.3281 & 0.7772 & 0.4616 \\
    10 & 0.8226 & 0.4554 & 0.8015 & 0.3018 & 0.7626 & 0.5064 & 0.7281 & 0.3902 & 0.7736 & \textbf{0.5054} & 0.7623 & 0.4602 \\
    \hline
  \end{tabular}
\end{table*}

\begin{table*}[t!]
  \centering
  \caption{Effectiveness of PU-based SS features when the number of topics are fixed for different classifiers.}
  \label{tab:effectiveness-PU-when-topic-fixed}
    \begin{tabular}{|c|c|c||c|c||c|c||c|c||c|c||c|c|}
    \hline
    \multirow{3}{*}{\#Words}
    &\multicolumn{2}{c||}{RF}
    &\multicolumn{2}{c||}{GBT}
    &\multicolumn{2}{c||}{MKL}
    &\multicolumn{2}{c||}{LR}
    &\multicolumn{2}{c||}{LSTM} 
    &\multicolumn{2}{c|}{Transformer}\cr 

    &\multicolumn{2}{c||}{(\#Topic=100)}
    &\multicolumn{2}{c||}{(\#Topic=60)}
    &\multicolumn{2}{c||}{( \#Topic=80)}
    &\multicolumn{2}{c||}{(\#Topic=80)}
    &\multicolumn{2}{c||}{(\#Topic=120)} 
    &\multicolumn{2}{c|}{(\#Topic=150)} \cr\cline{2-13} 

    &AUC &AP &AUC &AP &AUC &AP &AUC &AP &AUC &AP &AUC &AP \cr 

    \hline
    \hline
    50  &\textbf{0.9532} &\textbf{0.8238}&\textbf{0.9492}&\textbf{0.8008}  &0.9090     &0.5321          &0.6672         &0.2878 &0.8217 &\textbf{0.4753} &0.8119 &0.5813\cr\hline

    100 &0.9311          &0.8111               &0.9467          &0.7514         &0.9135     &\textbf{0.6966} &0.7054         &0.3665 &0.8014 &0.4458 &0.8551 &0.6048\cr\hline

    200 &0.9075          &0.6300              &0.9165          &0.7995         &\textbf{0.9224}    &0.5860          &\textbf{0.7778}&\textbf{0.5696} &0.8419 &0.4618 &0.8217 &0.5973\cr\hline

    400 &0.8693          &0.5964                &0.8869          &0.6529               &0.8468     &0.4327          &0.7466         &0.3964 &\textbf{0.8668} &0.4456 &0.8438 &\textbf{0.6152}\cr\hline

    600 &0.8547          &0.5836                  &0.8773          &0.6030                &0.8304     &0.4306          &0.5535         &0.1842 &0.7935 &0.4413 &\textbf{0.8565} &0.6025\cr\hline

    800 &0.8615          &0.5719                  &0.8572          &0.5388                &0.7860     &0.4664          &0.7360         &0.4859 &0.8169 &0.4517 &0.8388 &0.5914\cr\hline

    1000&0.8457          &0.6016               &0.8411          &0.5437               &0.8211     &0.4326          &0.7144         &0.2372 &0.8314 &0.4473 &0.8253 &0.5999\cr\hline
    \end{tabular}
\end{table*}

\subsubsection{The STL-based SS Feature}\label{sec:subsec:effectiveness-STL}

Table~\ref{tab:effectiveness-STL} shows the effectiveness of the number of Gaussian components on the STL-based SS feature. There are two observations from Table~\ref{tab:effectiveness-STL}:
\begin{itemize}
\item The optimal number of Gaussian components for the different classifiers is almost same for the neural network based method and the traditional machine learning method, respectively. For instance, 4 and 8 Gaussian components nearly obtain the optimal AUC and AP for ther neural networks and the traditional methods, respectively.
\item RF achieved the best performances in terms of AP. As expected, LSTM and Transformer  achieve a little less performance in terms of AP. Because the neural network usually needs more training data than these traditional methods.  
\end{itemize}

In summary, the optimal number of Gaussian components is as follows: 8 components for both RF and GBT, 10 components for MKL, and 4 components for both LSTM and Transformer.

\subsubsection{The PU-based SS Feature}\label{sec:subsec:effectiveness-PU}

Table~\ref{tab:effectiveness-PU-when-word-fixed} shows that the the GBT achieves the best performance (i.e., 0.9206 AUC and 0.8010 AP) among these baseline classifiers. Interestingly, RF and GBT are more
efficient than MKL in terms of AP on this feature. It means
that too more nonlinear kernels tend to over fit MKL classifier.

Moreover, different classifiers require a different number of topics to obtain a good performance. Compared with the STL-based SS features in Table~\ref{tab:effectiveness-STL}, the PU-based SS feature is more sensitive to the hyper-parameters. 
Once the optimal number of topics is determined for each classifier, Table~\ref{tab:effectiveness-PU-when-topic-fixed} shows the effectiveness of classifiers with respect to the number of words. The optimal number of words for different classifiers is slightly different. For instance the optimal number of words is 50, 50, 100, 200, 70, 100 for RF, GBT, MKL, LR, LSTM and Transformer respectively.

\begin{table}[t!]
  \centering
  \fontsize{7.5}{11}\selectfont
  \caption{Effectiveness of SS and pooling.}
  \label{tab:performace-SS-pooling}
    \begin{tabular}{|c|c|c||c|c|}
    \hline
    \multirow{2}{*}{Choices }
    &\multicolumn{2}{c||}{STL Behavior}
    &\multicolumn{2}{c|}{PU Behavior} \cr\cline{2-5}
    &AUC &AP &AUC &AP  \cr\hline\hline
    Without SS                  &\textbf{0.7480}   &0.4012            &0.7849           &0.4182  \cr\hline
    SS Without Pooling        &0.7266            &0.4258            &0.8047            &0.4257  \cr\hline
    SS With Pooling            &0.7391           &\textbf{0.4600}   &\textbf{0.8294}  &\textbf{0.4310}  \cr\hline
    \end{tabular}
\end{table}
In summary, the optimal number of words and the number of topics for the PU based features are (50,100), (50,60), (100,80), (200,80), (70,120) and (100,150) for RF, GBT, MKL, LR, LSTM and Transformer respectively. Note that without cross-validation on two hyper-parameters, we here determine the optimal parameters of the PU-based features by the alternative selection method, which is widely used to determine hyper-parameters~\cite{liu2012texture}.

\subsubsection{The gains of Self-Similarity and Pooling}

Table~\ref{tab:performace-SS-pooling} shows how SS and pooling affect the performances of classifiers:
 \begin{itemize}
 \item [1.]``\textbf{Without SS}'' means that either the STL behavior $\mathbf{f}^{STL}$ or the PU behavior $\mathbf{f}^{PU}$ is directly feeded into a classifier;
 \item [2.] ``\textbf{SS Without Pooling}'' means that the SS values in~\eqref{eqt:self-similarity} are firstly concatenated into a feature vector which is further feeded into a classifier;
\item [3.] ``\textbf{SS With Pooling}'' means the features in~\eqref{eqt:ss-based-feature} are feeded into a classifier.
\end{itemize}
RF is used as a baseline classifier. Table~\ref{tab:performace-SS-pooling} shows that the combination of SS and pooling significantly improves the performances of the behavior-based features. For instance, AUC and AP of the PU-based feature are improved from 0.7849 to 0.8294 and from 0.4182 to 0.4310, respectively.

For the STL-based feature, we also notice that the AUC of ``Without SS'' is reduced from 0.7480 to 0.7391; in contrast, the AP of ``SS With Pooling'' is improved from 0.4012 to 0.4600. Combining SS and pooling significantly increases the discriminative ability of the driver behaviors.

\begin{figure}[t!]
  \centering
  \subfigure[The duration time of sleeping]{
    \label{fig:subfig:t-d} 
    \includegraphics[width=.22\textwidth]{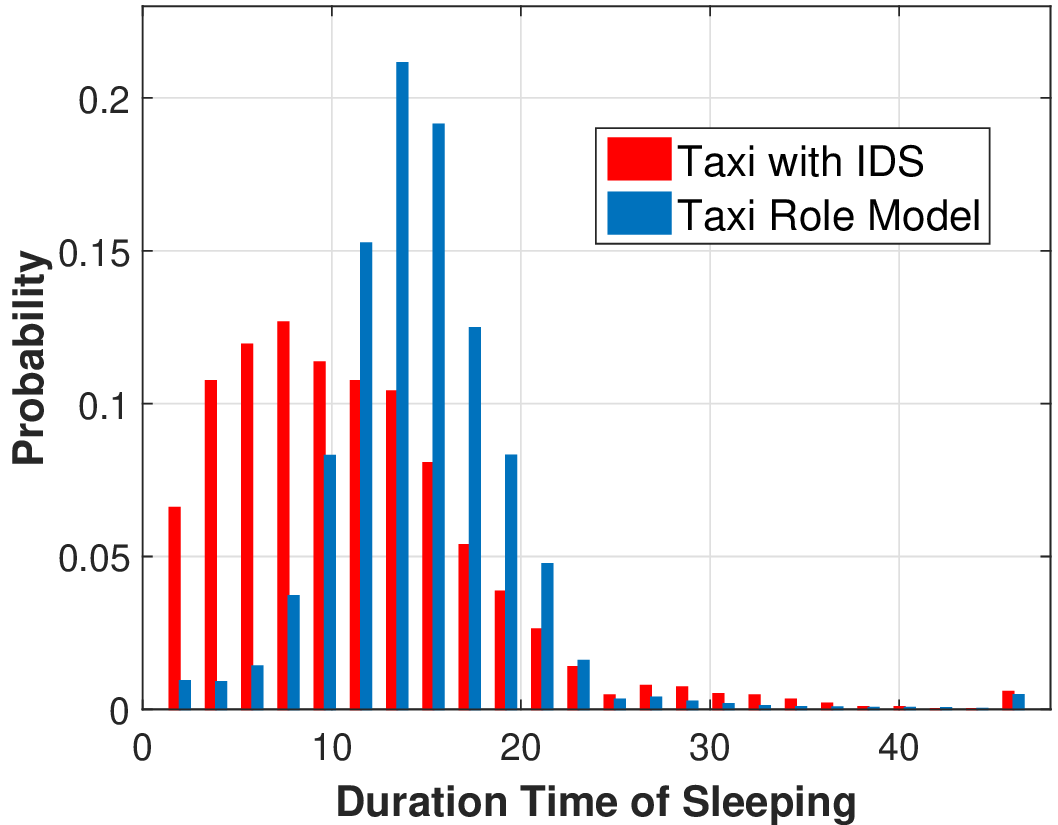}}
   \subfigure[The time when a taxi is started]{
    \label{fig:subfig:t-s} 
    \includegraphics[width=.23\textwidth]{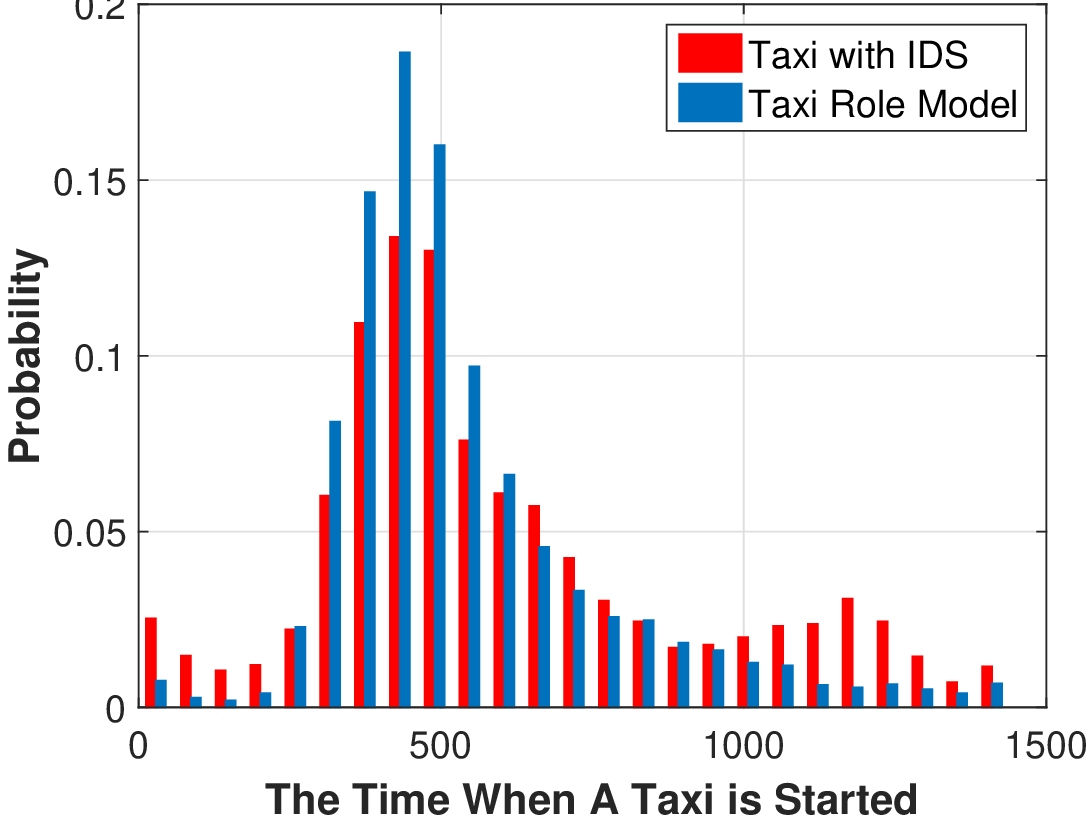}}
  \caption{The distributions of the two elements in STL.}
  \label{fig:distribtion-element-STL} 
\end{figure}

Table~\ref{tab:performace-SS-pooling} shows that ``Without SS'' slightly outperforms ``SS Without Pooling'' in terms of AUC, \emph{i.e.}, 0.7480 v.s. 0.7266. The explanation is that some elements in STL is very discriminative. To verify this intuition, Fig.~\ref{fig:distribtion-element-STL} simply shows the distributions of the time when a taxi is started $t_s$, and the duration time of sleeping $t_d$. As expected, the duration time of sleeping $t_d$ in Fig.~\ref{fig:subfig:t-s} is  discriminative in the behavior space to separate the positive samples and the negative ones. In contrast, the time when a taxi is started $t_s$ in Fig.~\ref{fig:subfig:t-d} barely contains any discriminative power in the behavior space.


\subsubsection{The combination of the STL-based and the PU-based SS features}

The AUC values from the combination of STL-based and PU-based features for each classifier are presented in Table.~\ref{tab:comparisons_baseline}. Note that, to obtain the best performance of different classifiers, the optimal parameters determined in Subsections~\ref{sec:subsec:effectiveness-STL} and~\ref{sec:subsec:effectiveness-PU} are adopted.

\begin{table}[t!]
\centering
\caption{Comparison of AUC Values Across Various Classifiers}
\label{tab:comparisons_baseline}
\begin{tabular}{@{}lcccccc@{}}
\toprule
& GBT & LR & MKL & RF & LSTM & Transformer \\ \midrule
AUC & 0.9472 & 0.8561 & 0.9488 &\textbf{0.9779} & 0.7487 & 0.6491 \\
\bottomrule
\end{tabular}
\end{table}

Table.~\ref{tab:comparisons_baseline} shows that GBT and RF achieve very similar performances and LSTM and Transformer classifier are relatively low. MKL slightly outperforms GBT in terms of AUC. The explanation is that GBT, RF, and MKL not only efficiently fuse multiple types of features but also have non-linear ability. Note that, to obtain the best performance of these different classifiers, the optimal parameters determined in Subsections~\ref{sec:subsec:effectiveness-STL} and~\ref{sec:subsec:effectiveness-PU} are adopted.

In summary, the empirical experiments show that the proposed behavior features have a good generalization ability across different classifiers.

\subsection{Effectiveness of MC-MIL}

In this subsection, we verify the effectiveness of the proposed MC-MIL by comparing with the following baseline methods:
\begin{itemize}
\item [1.] \textbf{Multiple Instance Boosting (MIL)~\cite{zhang2005multiple}:} MIL aligns the features by iteratively grouping a set of CARTs into a single strong learner. In this paper, the noise-or model in~\cite{zhang2005multiple} is used.
There are two categories of the hyper-parameters in MIL, i.e., the tree-specific parameters and the boosting parameters. In our setting, each instance is the concatenation of the SS-based STL~\eqref{eqt:fishervector} and the SS-based PU~\eqref{eqt:pickupfeature} in a $16$ days. That is, a bag ($30$ days) has 7 instances.
\item [2.] \textbf{Multiple Classifier Boosting (MCB)~\cite{kim2008mcboost}:} MCB aligns the samples into clusters in an iterative fashion. There are two categories of the hyper-parameters in MCL, the tree-specific parameters and the boosting parameters. These parameters are the same as in MIL. 
\end{itemize}

\subsubsection{Effectiveness of MC-MIL}

Tab.\ref{tab:effectiveness-MC-MIL} shows that MIL and MCL achieve very similar performances. As expected, MC-MIL outperforms both MIL and MCL in terms of AUC and AP. The explanation is that MC-MIL combines the advantages of the MIL, and MCL not only efficiently aligns the behaviors but also fuses multiple features. During training, the noise-or model in MCL requires that each classifier should be discriminative enough to obtain a good result.

In summary, the empirical experiments show that the proposed MC-MIL has a good generalization ability to handle both the alignment problem and  the feature fusion one.

\begin{table}[H]
  \centering
  \fontsize{7.5}{11}\selectfont
  \caption{Comparisons with two baseline methods.}
  \label{tab:effectiveness-MC-MIL}
    \begin{tabular}{|c|c||c|c||c|c|}
    \hline
    \multicolumn{2}{|c||}{MIL}
    &\multicolumn{2}{c||}{MCL}
    & \multicolumn{2}{c|}{MC-MIL} \cr\cline{1-6}
    AUC &AP &AUC &AP  & AUC & AP \cr\hline\hline
   0.8547   &0.7124  &0.8457  &0.7290 & \textbf{0.8937} &\textbf{0.7978}
     \cr\hline
    \end{tabular}
\end{table}

\subsection{Efficiency Evaluation}

Once the models for the STL behavior and the PU behavior are learned, the time complexity of encoding features per taxi would be linear with respect to the scales of samples. Table~\ref{tab:timecost-methods} shows that the STL-based SS feature and the PU-based SS feature consume 125.2 and 278.2 milliseconds per taxi, respectively. It means that the suspected taxis with IDS activity can be discovered within 7.5 hours on a single PC once the IDS model has been built. Note that the data set here was collected from all taxis in Beijing within 16 days; that is, the suspected taxis would be updated every 16 days.

Although the data sizes for discovering taxis with IDS may be prohibitively large for a single PC or a server, the proposed method can still be efficiently handled by incremental or online methods by the following methods:
\begin{itemize}
\item The GMM in FV can be efficiently solved by the online GMM~\cite{jaini2016online}. Once the GMM is learned, FV can be quickly encoded by~\eqref{eqt:FV-lambda} and~\eqref{eqt:FV-sigma}.
\item The online LDA~\cite{hoffman2010online} scales up the number of samples for all taxis.
\end{itemize}
Therefore, if these engineering details are properly implemented, the system would efficiently compute the suspected taxis with the IDS activity.

\begin{table}[t]
  \centering
  \caption{Offline Running times of different features and classifiers.}
  \label{tab:timecost-methods}
  \begin{tabular}{|c|c||c|c||c|c|}
    \hline
    \multicolumn{2}{|c||}{RF (Dep.=3)} & \multicolumn{2}{c||}{PU} & \multicolumn{2}{c|}{STL} \\\cline{1-6}
     Training & Testing & \multicolumn{1}{c|}{Behavior} & Encoding & \multicolumn{1}{c|}{Behavior} & Encoding \\
     (ms) & (ms) & \multicolumn{1}{c|}{Modeling} & by MSP & \multicolumn{1}{c|}{Modeling} & by MSP \\
     & & \multicolumn{1}{c|}{(s)} & (ms) & \multicolumn{1}{c|}{(s)} & (ms) \\
    \hline
  497.0 & 3.9 & 375.4 & 278.2 & 4.5183 & 125.2 \\
    \hline
     497.0 & 3.9 & 42.3 & 278.2 & 602.44 & 125.2 \\
    \hline
  \end{tabular}
  \footnotesize
  \begin{tablenotes}
    \item[*] ms. means millisecond, s. means seconds
  \end{tablenotes}
\end{table}

\section{Conclusions}\label{sec:conclusion}

This paper has described a computational approach to discovering one-shift taxis with the IDS activity. Based on the combination of the GPS traces and the records of taximeters, we propose a framework consisting of three phases: 1) modeling the STL behavior and the PU behavior for a taxi driver; 2) combining SS and MSP to map the behaviors into the IDS activity aligned feature space; 3) identifying the taxis with the IDS activity via the proposed MC-MIL classification. Extensive experiments are conducted on a real-life data set. The results demonstrate that the behavior-based features and the proposed MC-MIL achieve acceptable accuracy and efficiency.

In the future work, we aim to enhance our work as follows: 1) The limited number of PU behaviors indicates that the low-rank constraint may bring more efficient features than the application of LDA on the PU points; 2) We will attempt to construct a STL dictionary from the STL points, where each element in the dictionary represents a canonical user behavior; 3) how to build efficient driver behaviors for the two-shift taxis is an interesting direction; and 4) We will try to develop a semi-supervised method to reduce the number of the training data.

\bibliographystyle{IEEEtran}
\bibliography{refBus,refBus2,refBus1}
\vspace{-1.2cm}

\begin{IEEEbiography}[{\includegraphics[width=25mm,height=32mm,clip,keepaspectratio]{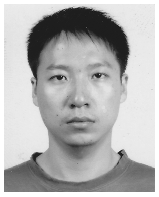}}]{Junbiao Pang}

received the B.S.degree and the M.S. degree in computational fluid dynamics and computer science
from the Harbin Institute of Technology, Harbin, China, in 2002 and 2004, respectively, and the
Ph.D. from the Institute of Computing Technology, Chinese Academy of Sciences, Beijing, China, in
2011.

He is currently an Associate Professor with Faculty of Information Technology, Beijing University of Technology (BJUT), Beijing, China. He has authored or coauthored approximately 20 academic papers in publications such as the IEEE TRANSACTIONS ON IMAGE
PROCESSING, ECCV, ICCV, and ACM Multimedia. His research interests include multimedia and machine learning for transportation problem.
\end{IEEEbiography}

\vspace{-1.2cm}

\begin{IEEEbiography}[{\includegraphics[width=25mm,height=32mm,clip,keepaspectratio]{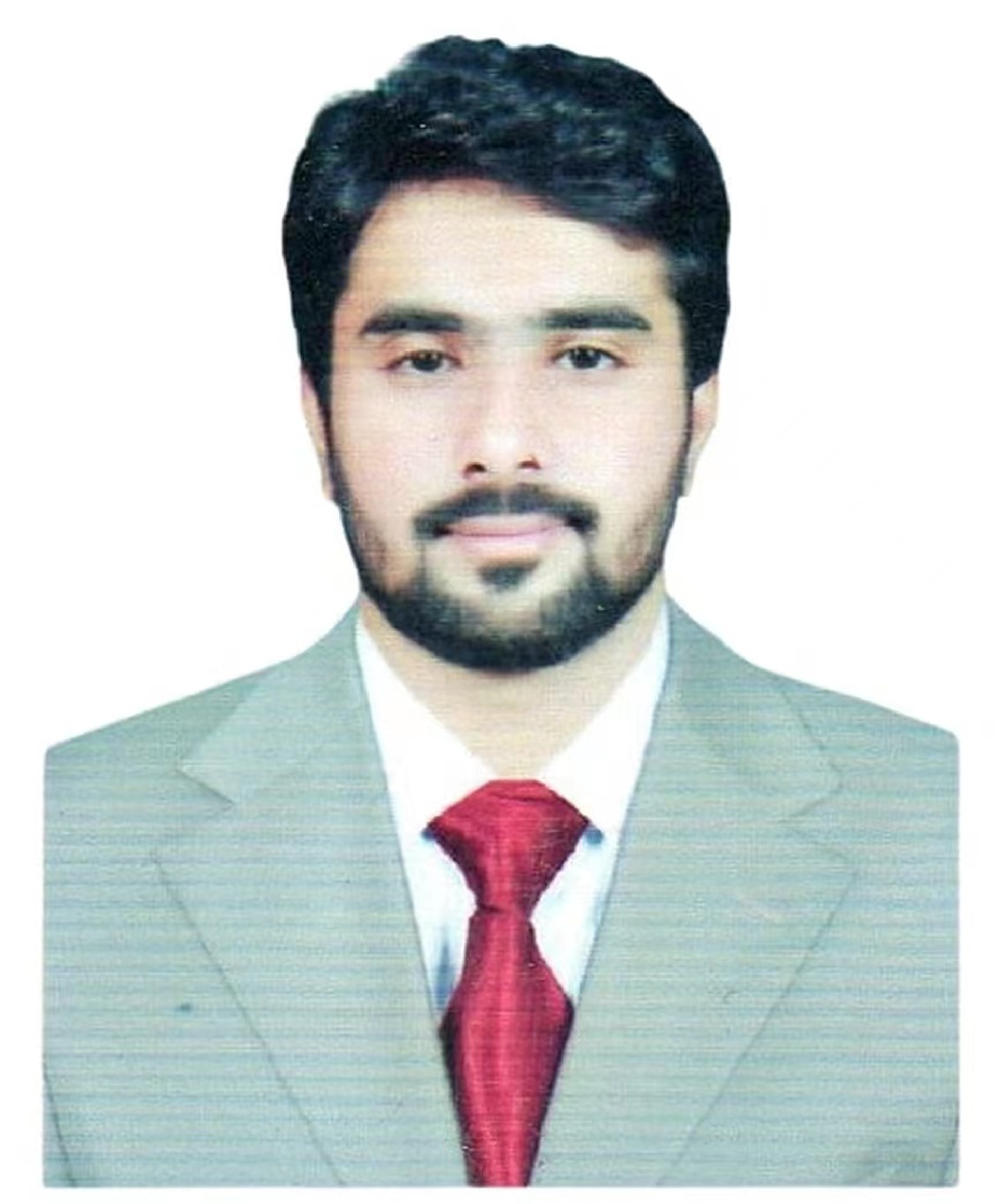}}]{Muhammad Ayub Sabir}

He earned his B.S. degree in Information Technology from the University of Sargodha, Pakistan, in 2017, and later completed his M.S. degree in the same field at Government College University Faisalabad, Pakistan, in 2019. Currently, he is pursuing a Ph.D. at Beijing University of Technology in the Department of Control Science and Engineering. His research interests span various areas, including Machine Learning, Computer Vision, and Image Processing.

\end{IEEEbiography}

\vspace{-1.2cm}

\begin{IEEEbiography}[{\includegraphics[width=25mm,height=32mm,clip,keepaspectratio]{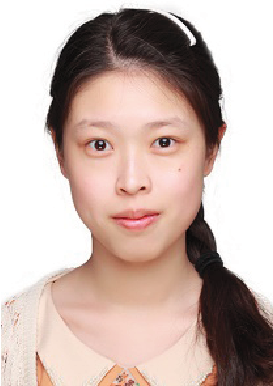}}]{Anjing Hu}

received the B.E. degree in electrical engineering and automation from Shanghai University, China, in 2016, and she is currently working towards
the M.S. degree in computer science and technology at Beijing University of Technology, Beijing, China. Her research interests include machine learning,
image content analysis, and information retrieval.
\end{IEEEbiography}

\vspace{-1.2cm}

\begin{IEEEbiography}[{\includegraphics[width=25mm,height=32mm,clip,keepaspectratio]{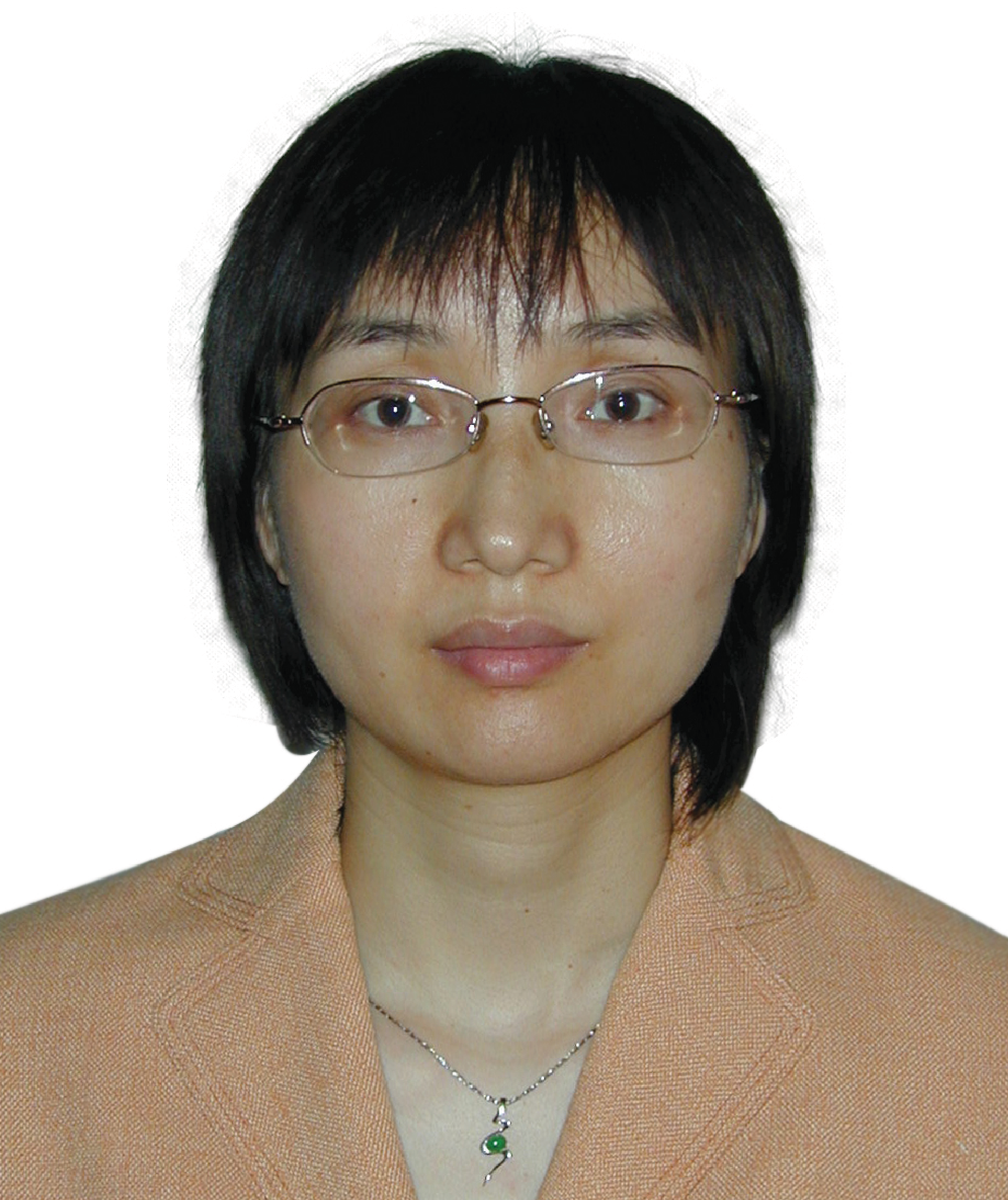}}]{Zuyun Wang}

received the B.S. degree in Software Engineer from the Capital Normal University, Beijing, China, in 1992, and the M.S. degree in Computer Application Technology from the Beijing University of Technology, Beijing, China, in 2003.

She is currently the director of the Information Center division of the Beijing Municipal Transportation Law Enforcement Corps, Beijing, China. Her research interests include the application of key technologies for intelligent transportation systems.
\end{IEEEbiography}

\vspace{-1.2cm}

\begin{IEEEbiography}[{\includegraphics[width=25mm,height=32mm,clip,keepaspectratio]{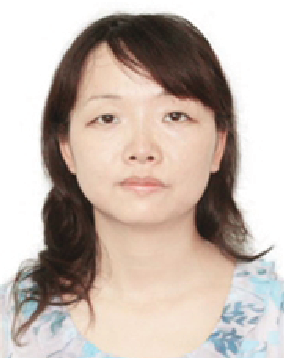}}]{Xue Yang}

received the B.S. degree and the M.S. degree in computer science and technology from the North China University of Water Resources and Electric Power, Zhengzhou, China, in 2001, and the Beijing University of Posts and Telecommunications, Beijing, China, in 2005, respectively.

She is currently a senior engineer at the Beijing Transportation Information Center, Beijing, China. Her research interests include traffic data analysis for the transportation problem.
\end{IEEEbiography}

\vspace{-1.2cm}

\begin{IEEEbiography}[{\includegraphics[width=25mm,height=32mm,clip,keepaspectratio]{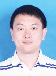}}]{Haitao Yu}

received the B.S. degree in management information systems from the Beijing Information Science and Technology University, Beijing, China, in 2005, and the M.S. degree in computer software from the Beihang University, Beijing, China, in 2009, and is currently working towards the Ph.D. degree in computer science and technology at Beihang University, Beijing, China.
He is currently the associate dean with the Beijing Transportation Information Center, Beijing, China. His research interests include traffic data analysis for the key technologies in traffic information service.
\end{IEEEbiography}

\vspace{-1.2cm}

\begin{IEEEbiography}[{\includegraphics[width=25mm,height=32mm,clip,keepaspectratio]{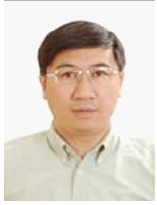}}]{Qingming Huang}

( M'04 - SM'08) received the B.S. degree in computer science and the Ph.D. degree in computer engineering from Harbin Institute
of Technology, Harbin, China, in 1988 and 1994, respectively.

He is currently a Professor with the University of the Chinese Academy of Sciences (CAS), Beijing, China, and an Adjunct Professor with the Institute of Computing Technology, CAS, China. He has authored or coauthored more than 300 academic papers in prestigious international journals including the IEEE TRANSACTIONS ON MULTIMEDIA, and the IEEE TRANSACTIONS ON IMAGE PROCESSING, and top-level conferences such as ACM Multimedia, CVPR, AAAI, IJCAI and VLDB. His research interests include multimedia computing, image processing, computer vision, pattern recognition, and machine learning.

\end{IEEEbiography}

\end{document}